\documentclass[camera,letterpaper,nomarginnotes,nonarrowgutter]{jpaper}

\newif\ifdraft
\draftfalse
\newif\ifrebuttal
\rebuttalfalse
\newif\ifcameraready
\camerareadytrue
\newif\ifarXiv
\newif\ifacmversion
\acmversionfalse
\newif\ifpagenumbers

\newcommand{\mitigatingRowHammerAllCitations}[0]{\cite{AppleRefInc, rh-hp,rh-lenovo,greenfield2012throttling, kim2014flipping, kim2014architectural, bains14d, bains14c, bains14a, bains14b, aweke2016anvil, bains2015row, bains2016row, seyedzadeh2016counter, bains2016distributed, greenfield2016throttling, son2017making, seyedzadeh2018cbt,irazoqui2016mascat, you2019mrloc, lee2019twice, park2020graphene, yaglikci2021security, yaglikci2021blockhammer, frigo2020trrespass, kang2020cattwo, hassan2021utrr, qureshi2022hydra, saileshwar2022randomized, brasser2017can, konoth2018zebram, van2018guardion, vig2018rapid,  kim2022mithril, lee2021cryoguard, marazzi2022protrr, zhang2022softtrr, joardar2022learning, juffinger2023csi, yaglikci2022hira, saxena2022aqua, enomoto2022efficient, manzhosov2022revisiting, ajorpaz2022evax, naseredini2022alarm, joardar2022machine, tomita2022extracting, zhang2020leveraging,loughlin2021stop, devaux2021method, fakhrzadehgan2022safeguard, saroiu2022price, loughlin2022moesiprime, han2021surround, mutlu2023fundamentally, woo2023scalable, marazzi2023rega, bock2019riprh, wang2021discreet, bennett2021panopticon, olgun2024abacus, bostanci2024comet, canpolat2024understanding, saxena2024start, saxena2024rubix, qureshi2024mint, saxena2024impress, kim20231ddr5, canpolat2025chronus, canpolat2024breakhammer, tugrul2025understanding,yaglikci2024svard,taneja2025dream,vittal2025mopac,lin2025cnc,qazi2025drfm,lu2025counterpoint,qureshi2025autorfm,woo2025dapper,woo2025qprac,bostanci2025understanding}}

\usepackage{amsmath,amssymb,amsfonts}
\usepackage{graphicx}
\usepackage{textcomp}
\usepackage{url}
\usepackage{algorithm}
\usepackage{algpseudocode}
\usepackage{multirow}
\usepackage{tikz}
\usepackage{marginnote}
\usepackage{fancyhdr}
\usepackage{titlesec}
\usepackage{dblfloatfix}
\usepackage{xspace}
\usepackage{enumitem}
\usepackage{xcolor}
\usepackage{booktabs}
\usepackage{float}
\usepackage{caption}
\usepackage[most]{tcolorbox}
\usepackage{siunitx}
\usepackage{setspace}
\usepackage{microtype}
\usepackage{fourier-orns}
\usepackage{threeparttable}
\usepackage{balance}
\usepackage{mathptmx}

\usepackage{yfonts}
\usetikzlibrary{calc}
\usepackage{pgfornament}

\PassOptionsToPackage{bookmarks=true,breaklinks=true,colorlinks,citecolor=blue,linkcolor=blue,urlcolor=blue}{hyperref}
\usepackage{hyperref}
\usepackage[sort,compress]{cite}
\usepackage{flafter}
\usepackage{pdflscape}
\usepackage{makecell}
\usepackage{siunitx}
\usepackage{multibib}
\usepackage{xurl}

\definecolor{gfored}{rgb}{0.580, 0.050, 0.211}
\definecolor{ao}{rgb}{0.007, 0.520, 0.867}
\definecolor{moegi}{rgb}{0.357, 0.537, 0.188}
\definecolor{jl}{rgb}{1.0, 0.2, 0.8}
\definecolor{brown(web)}{rgb}{0.65, 0.16, 0.16}
\definecolor{bisque}{rgb}{1.0, 0.89, 0.77}
\definecolor{nbs}{rgb}{0.88, 0.07, 0.37}
\definecolor{yt}{rgb}{0.58, 0.44, 0.86}
\definecolor{iy}{rgb}{0.0, 0.36, 0.05}
\definecolor{mel}{rgb}{0.9, 0.55, 0.31}
\definecolor{ouscolor}{rgb}{0.0, 0.2, 0.4}
\definecolor{jwangc}{rgb}{0.88, 0.07, 0.37}

\definecolor{cminor}{rgb}{0.16, 0.35, 0.63}     
\definecolor{ccommon}{rgb}{0.0, 0.47, 0.43}     
\definecolor{cindiv}{rgb}{0.35, 0.20, 0.59}     

\ifarXiv
    \newcommand{\jwarv}[2]{\ifnum#1>-1\textcolor{red}{#2}\else{#2}\fi}
\else
    \newcommand{\jwarv}[2]{\ifnum#1>-1\textcolor{black}{#2}\else{#2}\fi}
\fi
\ifcameraready
    \newcommand{\jwcr}[2]{\ifnum#1>-1\textcolor{black}{#2}\else{#2}\fi}
    \newcommand{\atbcr}[2]{\ifnum#1>-1\textcolor{black}{#2}\else{#2}\fi}
    \newcommand{\hluocr}[2]{\ifnum#1>-1\textcolor{black}{#2}\else{#2}\fi}
    \newcommand{\ieycr}[2]{\ifnum#1>-1\textcolor{black}{#2}\else{#2}\fi}
    \newcommand{\omcr}[2]{\ifnum#1>-1\textcolor{black}{#2}\else{#2}\fi}
    \newcommand{\omcrcomment}[1]{}
    \newcommand{\crdiscussion}[2]{}{}

    \newcommand{\ominline}[1]{}
    \newcommand{\jwcrcomment}[1]{}
    \newcommand{\ieycrcomment}[1]{}
    \newcommand{\atbcrcomment}[1]{}
    \newcommand{\agycrcomment}[1]{}
    \newcommand{\ieyinline}[1]{}
\else
\usepackage[colorinlistoftodos,prependcaption,textsize=scriptsize]{todonotes}
    \paperwidth=\dimexpr \paperwidth + 4cm\relax
    \oddsidemargin=\dimexpr\oddsidemargin + 2cm\relax
    \evensidemargin=\dimexpr\evensidemargin + 2cm\relax
    \marginparwidth=\dimexpr \marginparwidth + 2cm\relax

    \newcommand{\jwcr}[2]{\ifnum#1=\value{version}\textcolor{blue}{#2}\else{#2}\fi}
    \newcommand{\jwcrcomment}[1]{\todo[linecolor=brown, bordercolor=brown, backgroundcolor=white]{\textcolor{jwangc}{\textbf{Jikun:} #1}}}

    \newcommand{\atbcr}[2]{\ifnum#1=\value{version}\textcolor{ao}{#2}\else{#2}\fi}

    \newcommand{\ieycr}[2]{\ifnum#1=\value{version}\textcolor{blue}{#2}\else{#2}\fi}

    \newcommand{\ieycrcomment}[1]{\todo[linecolor=orange, bordercolor=orange, backgroundcolor=white]{\textcolor{iy}{\textbf{@Ismail:} #1}}}
    \newcommand{\ieyinline}[1]{\\\textcolor{iy}{\textbf{@Ismail:} #1}}

    \newcommand{\atbcrcomment}[1]{\todo[linecolor=brown, bordercolor=brown, backgroundcolor=white]{\textcolor{ao}{\textbf{@Atb:} #1}}}

    \newcommand{\agycrcomment}[1]{\todo[size=\scriptsize, linecolor=orange, bordercolor=orange, backgroundcolor=white]{\textcolor{gfored}{\textbf{@gy:} #1}}}

    \newcommand{\crdiscussion}[2]{\omcrcomment{#1\\\textcolor{blue}{\textbf{@Jikun:}#2}}}
    \newcommand{\omcr}[2]{\ifnum#1=\value{version}\textcolor{red}{#2}\else{#2}\fi}

    \newcommand{\omcrcomment}[1]{\todo[linecolor=orange, bordercolor=orange, backgroundcolor=white]{\textcolor{red}{\textbf{@Onur:} #1}}}
    \newcommand{\ominline}[1]{\\\textcolor{red}{\textbf{@Onur:} #1}}
    
    \newcommand{\hluocr}[2]{\ifnum#1=\value{version}\textcolor{moegi}{#2}\else{#2}\fi}
\fi

\ifrebuttal
    \newcommand{\rbm}[1]{\textcolor{blue!70!black}{#1}}
    \newcommand{\rbi}[1]{\textcolor{orange!85!black}{#1}}
    \newcommand{\rbc}[1]{\textcolor{purple!90!black}{#1}}
    
    \newcommand{\concern}[1]{\colorbox{yellow!40}{\hyperref[ref:#1]{{#1}}}}
    \newcommand{\cref}[1]{\hyperref[ref:#1]{#1}}
    \newcommand\revmid[2]{\todo[size=\scriptsize, linecolor=#1,backgroundcolor=#1!15,bordercolor=#1]{\textcolor{black}{\textbf{#2}}}}

    \newcommand{\revci}[1]{\revmid{orange}{#1}\label{ref:#1}}
    \newcommand{\revcc}[1]{\revmid{purple!90!white}{#1}\label{ref:#1}}
    
\else
    \newcommand{\rbm}[1]{{#1}}
    \newcommand{\rbc}[1]{\textcolor{black}{#1}}
    \newcommand{\rbi}[1]{\textcolor{black}{#1}}

    \newcommand{\revcc}[1]{}
    \newcommand{\revci}[1]{}
    \newcommand{\revcm}[1]{}
\fi

\ifdraft
\usepackage[colorinlistoftodos,prependcaption,textsize=scriptsize]{todonotes}
    \newcommand{\hluo}[1]{\textcolor{black}{#1}}
    \newcommand{\atb}[1]{\textcolor{black}{#1}}
    \newcommand{\atbcomment}[1]{#1}
    \newcommand\nb[1]{{\color{black}{#1}}}
    \newcommand{\nbcomment}[1]{\todo[size=\scriptsize, linecolor=orange, bordercolor=orange, backgroundcolor=white]{\textcolor{nbs}{NB:~#1}}}
    \newcommand\iey[1]{{\color{black}{#1}}}
    \newcommand{\agycomment}[1]{#1}

    \newcommand{\jwcomment}[1]{\todo[size=\scriptsize, linecolor=orange, bordercolor=orange, backgroundcolor=white]{\textcolor{jwangc}{Jikun:~#1}}}
\else
    \newcommand{\hluo}[1]{\textcolor{black}{#1}}
    \newcommand{\atb}[1]{\textcolor{black}{#1}}
    \newcommand{\atbcomment}[1]{}
    \newcommand\nb[1]{{\color{black}{#1}}}
    \newcommand{\nbcomment}[1]{}
    \newcommand\iey[1]{{\color{black}{#1}}}
    \newcommand{\agycomment}[1]{}

    \newcommand{\jwcomment}[1]{}
\fi

\newcommand\proposal{\jwcr{1}{{ScaleDisturb}}\xspace}
\newcommand\chips{196\xspace}

\newcommand\threshold{\emph{ACmin}\xspace}
\newcommand\sol{\emph{TeACUp}\xspace}
\newcommand{\secref}[1]{§\ref{#1}}
\newcommand{\figref}[1]{Fig.~\ref{#1}}

\newcommand{\tRAS}{$t_{\textit{RAS}}$}
\newcommand\tRP{$t_{\textit{RP}}$}
\newcommand{\tRC}{$t_{\textit{RC}}$}
\newcommand\tagg{$t_{\textit{AggON}}$}
\newcommand\taggone{$t_{\textit{AggON1}}$}
\newcommand\taggtwo{$t_{\textit{AggON2}}$}

\newcommand\baseline{double-sided RowPress\xspace}

\newcommand\rowagx{$\textit{R}_{N\!-\!1}$}
\newcommand\rowagy{$\textit{R}_{N\!+\!1}$}
\newcommand\rowv{$\textit{R}_{N}$}

\newcommand\meanavr{9.6\%\xspace}
\newcommand\maxavr{63\%\xspace}
\newcommand\meanmvr{16.1\%\xspace}
\newcommand\maxmvr{52.4\%\xspace}

\newtcolorbox{obsvboxstyle}{%
    enhanced,
    colback=black!5,
    colframe=black!75, 
  boxrule=0.8pt,       
  arc=0.8pt,             
  left=1pt, right=1pt, top=0.5pt, bottom=0.5pt,
  before skip=5pt,
  after skip=5pt,
}

\newcounter{observationctr}
\newenvironment{obsvbox}
{
  \stepcounter{observationctr}
  \begin{obsvboxstyle}
  \textbf{Observation~\theobservationctr.}%
}
{
  \end{obsvboxstyle}
}

\newtcolorbox{takeawayboxstyle}{
  enhanced,
  colback=brown!10,
  colframe=brown!50,      
  boxrule=1pt,       
  arc=1pt,             
  left=1pt, right=1pt, top=0.5pt, bottom=0.5pt,
  before skip=4pt,
  after skip=4pt,
}

\newcounter{takeawayctr}
\newenvironment{takeawaybox}
{
  \stepcounter{takeawayctr}
  \begin{takeawayboxstyle}
  \textbf{Takeaway~\thetakeawayctr.}%
}
{
  \end{takeawayboxstyle}
}

\lstdefinestyle{cppStyle}{
  language=C++,
  basicstyle=\ttfamily\scriptsize,
  keywordstyle=\color{blue}\bfseries,
  commentstyle=\color{gray}\itshape,
  stringstyle=\color{teal},
  numbers=left,
  numberstyle=\tiny,
  numbersep=8pt,
  frame=single,
  breaklines=false,
  showstringspaces=false,
  xleftmargin=0.5em,
  framexleftmargin=0.5em,
  xrightmargin=0.5em,
  tabsize=2,
  morekeywords=[2]{initialize, clflushopt, mfence, activate_dummy_rows, check_bitflips, find_aggressor_rows},
  keywordstyle=[2]\color{blue},
  morekeywords=[3]{NUM_READ1, NUM_READ2},
  keywordstyle=[3]\color{red}
}

\newcommand{\circledcharblack}[1]{%
    \tikz[baseline=(char.base)]{
        \node[shape=circle, 
        minimum size=8pt,
        inner sep=0pt,
        draw=black, fill=black, line width=0.5pt] (char) {\textcolor{white}{\textbf{#1}}};
    }
}

 \makeatletter
 \g@addto@macro{\normalsize}{%
  \setlength{\abovedisplayskip}{2pt plus 1pt minus 1pt}
  \setlength{\belowdisplayskip}{2pt plus 1pt minus 1pt}
   \setlength{\abovedisplayshortskip}{0pt}
   \setlength{\belowdisplayshortskip}{0pt}
   \setlength{\intextsep}{2pt plus 1pt minus 1pt}
   \setlength{\textfloatsep}{2pt plus 1pt minus 1pt}
   \setlength{\skip\footins}{5pt plus 1pt minus 1pt}}
 \makeatother

\newcites{mod}{Module References}
\newcommand{\param}[1]{\textcolor{black}{#1}}

\newcounter{version}
\begin{document}
\title{{\proposal: Exploiting Temporal Asymmetry \\to Amplify Read Disturbance in Modern DRAM Chips}}

\newcommand{\affilETH}[0]{\textsuperscript{\S}}
\newcommand{\affilCISPA}[0]{\textsuperscript{$\dagger$}}
\newcommand{\affilINRIA}[0]{\textsuperscript{$^\ddagger$}}
\author{
{Jikun Wang\affilETH}\hspace{1.2em}
{Haocong Luo\affilETH}\hspace{1.2em}
{Ataberk Olgun\affilETH}\hspace{1.2em}
{{\.I}smail Emir Y{\"u}ksel\affilETH}\hspace{1.2em}
{A.~Giray~Ya\u{g}l{\i}k\c{c}{\i}\affilETH\textsuperscript{,}\affilCISPA}\hspace{1.2em}\\
{Yu Liang\affilINRIA}\hspace{1.2em}
{F. Nisa Bostanc{\i}\affilETH}\hspace{1.2em}
{Mohammad Sadrosadati\affilETH}\hspace{1.2em}
{Onur Mutlu\affilETH}
\\
{\affilETH \emph{ETH Z{\"u}rich}\hspace{1.2em} \affilCISPA \emph{CISPA}\hspace{1.2em} \affilINRIA \emph{Inria Paris}}
}

\maketitle

\fancypagestyle{firstpage}
{
    \fancyhead{}
    \begin{tikzpicture}[remember picture,overlay]
    \node [xshift=150mm,yshift=-10mm]
    at (current page.north west) {{\includegraphics[width=3cm]{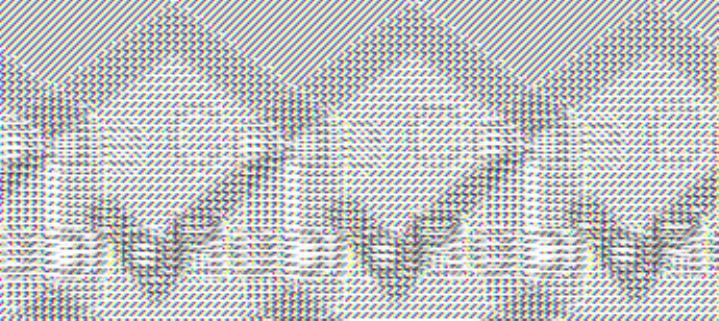}}} ;
    \node [xshift=183mm,yshift=-10mm]
    at (current page.north west) {{\includegraphics[width=3cm]{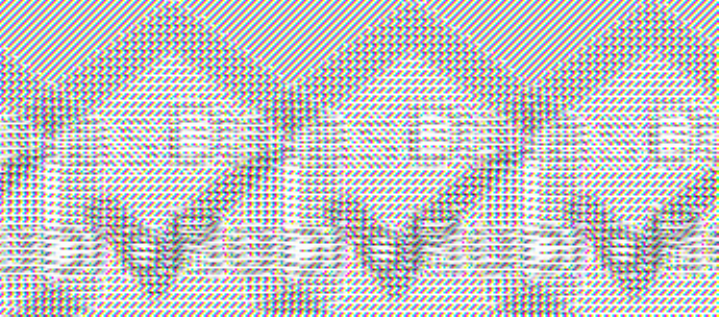}}} ;
    \end{tikzpicture}
  \renewcommand{\headrulewidth}{0pt}
}
\thispagestyle{firstpage}
\pagenumbering{arabic}

\setcounter{version}{4}
\begin{abstract}

\omcr{2}{DRAM suffers} from read disturbance phenomena (e.g., RowHammer and RowPress), where repeatedly accessing \omcr{2}{or continuously keeping open} a DRAM row (aggressor row) induces bitflips in other physically nearby \emph{unaccessed} rows (victim rows). The disturbance mechanism is practically exploitable from the software stack and worsens across generations \jwcr{1}{with continued density scaling}. DRAM read disturbance is highly sensitive to memory access patterns, yet prior work \jwcr{1}{explores read disturbance under only a \emph{limited} set of access patterns.}

~\omcr{3}{W}e present \jwcr{1}{\proposal, a new DRAM access pattern that can amplify DRAM} read disturbance by asymmetrically extending the open time of two aggressor rows. \jwcr{1}{Our} rigorous experimental characterization of 196 DDR4 and 3 HBM2 DRAM \jwcr{1}{chips shows} that \proposal 
(1)~\omcr{2}{leads to} bitflips at significantly fewer row activations, compared \rbm{to state-of-the-art} memory access patterns,
(2)~makes read disturbance attacks easier across all tested DRAM chips,
(3)~\omcr{2}{increases DRAM vulnerability to read disturbance} as DRAM manufacturing technology scales down~\omcr{3}{to smaller node sizes}.
We showcase a proof-of-concept attack on a real system where a user-level program \omcr{2}{leveraging \proposal} induces more bitflips \omcr{2}{than} \rbm{state}-of-the-art~\omcr{2}{RowHammer and RowPress} memory access patterns.
\omcr{2}{We describe and evaluate four solutions for mitigating read disturbance bitflips in the presence of \proposal and call for more research on the topic.}

\end{abstract}
\setcounter{version}{4}


\section{Introduction}
\label{sec:Intro}
Memory isolation is \jwcr{1}{fundamental to} system robustness\jwcr{1}{, including safety, security, reliability, \omcr{2}{and availability}}. Accesses to one memory location should not induce unintended \jwcr{1}{side effects} on other unaccessed memory locations. Unfortunately, \jwcr{1}{dynamic random access memory (DRAM)\omcr{2}{\cite{dennard1968field,mutlu2013memory,mutlu2025memory}}, the dominant main memory technology}, is vulnerable to \emph{read disturbance}\rbm{, whereby} accessing \jwcr{1}{one} DRAM cell \jwcr{1}{can disturb} the integrity of data in physically adjacent \jwcr{1}{yet unaccessed cells}, \rbm{thereby violating} memory isolation\omcr{2}{~\cite{kim2014flipping,mutlu2017rowhammer,mutlu2019rowhammer,kim2020revisiting,frigo2020trrespass,orosa2021deeper,hassan2021utrr,kogler2022half,luo2023rowpress,luo2024rowpress,mutlu2023fundamentally,luo2024experimental,luo2025revisiting}}.

\hluo{Prior works identify two widespread DRAM read disturbance phenomena,} RowHammer\omcr{2}{~\cite{kim2014flipping,mutlu2017rowhammer,mutlu2019rowhammer, kim2020revisiting, frigo2020trrespass, orosa2021deeper, hassan2021utrr, kogler2022half, mutlu2023fundamentally, olgun2025vrd,yaglikci2024svard,luo2024experimental, luo2025revisiting} }and RowPress\omcr{2}{~\cite{luo2023rowpress,luo2024rowpress}}\hluo{, in modern DRAM chips. RowHammer} \jwcr{1}{induces} bitflips in a DRAM row (\emph{victim row}) by repeatedly opening and closing a physically adjacent DRAM row (\emph{aggressor row}). RowPress \jwcr{1}{induces} bitflips in the victim row by keeping the aggressor row open for \jwcr{1}{an extended} period of time (i.e., aggressor row on time, \emph{\tagg{}}).

\jwcr{1}{DRAM chips' vulnerability} to read disturbance \jwcr{1}{strongly depends on} DRAM row access patterns. 
For example, prior works~\cite{kim2014flipping, kim2020revisiting, orosa2021deeper, olgun2024hbm,lang2023blaster} show that a double-sided RowHammer \nb{access} pattern \jwcr{1}{(i.e., alternately activating two adjacent aggressor rows \omcr{2}{within an access loop})} is much more effective (i.e., requir\omcr{2}{es} \jwcr{1}{far} fewer aggressor row activations to induce bitflips) than a single-sided RowHammer \nb{access} pattern, \omcr{2}{which activates only one aggressor row \omcr{2}{in each access loop}}. 
For RowPress, a longer \tagg{} makes the RowPress access pattern more effective at inducing bitflips \jwcr{1}{than RowHammer access patterns}\jwarv{2}{~\cite{orosa2021deeper,luo2023rowpress, luo2024rowpress, luo2024experimental,yaglikci2024svard,olgun2024hbm,olgun2023hbm,luo2025revisiting,olgun2025vrd}}. 
\jwcr{1}{Prior characterization works conservatively quantify a DRAM row's read disturbance vulnerability under the most effective access pattern using a row-level metric: the minimum number of aggressor row activations required to induce \omcr{2}{a bitflip} in a victim row (\threshold). 
The reported minimum \threshold across all tested rows in a DRAM module is then used as the threshold for triggering protective operations in read disturbance mitigation mechanisms to \omcr{2}{prevent} any erroneous bits. \omcr{2}{I}dentifying more effective DRAM row access patterns is critical to \omcr{2}{1)} accurately characteriz\omcr{2}{ing} DRAM vulnerability and \omcr{2}{2)} design\omcr{2}{ing} robust mitigation mechanisms.}
However, \rbm{prior work\omcr{2}{s} cumulatively examine subset of all possible memory access scenarios.}

\textbf{Our goal} in this paper is to explore a \omcr{2}{new} DRAM access \rbm{pattern} that can reduce the minimum aggressor row activation count \jwcr{1}{required} to induce bitflips compared to prior access patterns, and \jwcr{1}{to evaluate} \rbm{its} implications on modern DRAM-based systems.
\jwcr{1}{To this end,} we introduce \emph{\proposal}, a new memory access pattern that (i) repeatedly takes turns \rbm{activating} two aggressor rows~\atb{(\rowagx~and~\rowagy)}~\atb{physically adjacent to a victim row (\rowv),} and (ii)~\jwcr{1}{keeps} the two aggressor rows open for \emph{different} \rbm{durations, thereby enabling temporal asymmetry that distinguishes it from existing double-sided patterns \omcr{2}{that use} symmetric aggressor row open times}. 

We characterize \proposal on commercial off-the-shelf (COTS) DRAM chips from three major DRAM manufacturers, including~\chips DDR4 \jwcr{1}{chips spanning} 15 chip revisions and 3 HBM2 chips.
Our results show that, \omcr{2}{under the same extra row \emph{open time budget} (OTB) assigned to the two aggressor rows in each access loop, \proposal requires fewer aggressor row activations to induce bitflips than double-sided RowPress.}
Across all individual victim rows we test, \proposal reduces \threshold by \meanavr on average (up to \maxavr~when \omcr{2}{the OTB is large}). For all DRAM modules we test, \proposal reduces the \threshold of a module (i.e., the minimum \threshold across all victim rows in a module) by \meanmvr~on average (up to \maxmvr when \omcr{2}{the OTB is large}).

Our detailed characterization \omcr{2}{reveals} four key findings. First, \proposal~\nb{is} widespread across all tested DDR4 and HBM2 DRAM chips from all three major manufacturers and worsens as DRAM technology scales down to smaller node sizes. Second, \proposal reduce\nb{s} \threshold in a non-negligible fraction (37.5\% on average, up to 49.2\%) of the victim DRAM rows we test. Third, \proposal's \threshold reduction is stable over time and is not attributable to Variable Read Disturbance (VRD)~\cite{olgun2025vrd}. Fourth, \proposal flips a different set of bits compared to double-sided RowPress\omcr{2}{~\cite{luo2023rowpress,luo2024rowpress}}. 
We hope our findings enable future works to advance the understanding of DRAM read disturbance under different aggressor row access patterns.

We demonstrate on a real DDR4-based system \jwcr{1}{equipped} with an in-DRAM RowHammer mitigation mechanism\omcr{2}{~\cite{hassan2021utrr}} that a user-level program \rbm{that uses} \proposal i) induce\omcr{2}{s} bitflips in scenarios where \nb{conventional} double-sided RowPress \nb{cannot} and ii) induce\omcr{2}{s} more bitflips than \jwcr{1}{double-sided} RowPress in certain scenarios (i.e., when the program accesses dummy rows and synchronizes with DRAM auto-refresh \emph{less frequently}).

Our results have implications for the security guarantees of read disturbance mitigation mechanisms\omcr{2}{, as they rely on accurate \omcr{3}{identification of} \threshold to configure protection thresholds. I}f the \threshold of a DRAM row is not identified accurately~\jwcr{1}{(i.e., overestimated when characterized using existing access patterns)}, these mechanisms can easily become insecure. 
We evaluate and \jwcr{1}{examine} three potential solutions for mitigating read disturbance bitflips in the presence of \proposal.
\jwcr{1}{Our results show that (i) the number of bitflips induced by \proposal far exceeds \omcr{2}{simple} ECC's correction capability, and (ii) applying safety margins to state-of-the-art mitigation mechanisms incurs significant performance overhead (\param{8.8\%} on average, up to \param{28.6\%}) and energy overhead (\param{12.3\%} on average, up to \param{58.8\%}).}

We propose \omcr{2}{\emph{Temporal Asymmetry-aware Counter Update} (\emph{TeACUp})}, a new mitigation mechanism that effectively mitigates \proposal with low \emph{additional} overhead. 
The key idea of \emph{TeACUp} is to apply a \emph{dynamic scal\omcr{2}{ing} ratio} \omcr{2}{to mitigate the asymmetric growth of row activation counters introduced by \proposal},
\rbm{preventing \omcr{2}{one-sided counters} from prematurely reaching the \omcr{2}{protection} threshold.} 

\noindent We make the following contributions:
\begin{itemize}[leftmargin=*]
    \item To our knowledge, this is the first work to experimentally demonstrate that keeping the two aggressor rows open for \omcr{2}{\emph{different}} amounts of time \omcr{2}{(i.e., \omcr{3}{the} \proposal access pattern)} induce\omcr{2}{s} bitflips in the victim row with fewer aggressor row activations \jwcr{1}{than} existing DRAM access patterns characterized in prior works.
    
    \item We provide an extensive characterization of \proposal on \omcr{3}{\chips{} real} DDR4 and 3 HBM2 DRAM chips. Our analys\rbm{e}s show that~\proposal(1)~significantly amplif\rbm{ies} DRAM \jwcr{1}{chips'} vulnerability to read disturbance, (2)~widely affects DRAM chips from three \jwcr{1}{major DRAM} manufacturers, \omcr{2}{(3)} worsens as DRAM \jwcr{1}{manufacturing} technology scales down, (\omcr{2}{4})~fundamentally differs from Variable Read Disturbance in \rbm{reproducibility}, stability and \omcr{2}{magnitude of \threshold reduction}, and (\omcr{2}{5})~flips a different set of bits compared to double-sided RowPress.
    
    \item We demonstrate on a real system with RowHammer protection that a user-level program leveraging \proposal induce\omcr{2}{s} \omcr{2}{(1)} bitflips \omcr{2}{where} RowPress \rbm{cannot, and} \omcr{2}{(2) induces more bitflips \omcr{3}{than RowHammer and RowPress}}.
    
    \item We evaluate and \jwcr{1}{examine} three \omcr{2}{classes of} potential solutions \rbm{to mitigate} \proposal bitflips.~We propose \sol, a new read disturbance mitigation mechanism that effectively mitigates \proposal with low additional overhead.

    \item \omcr{2}{We call for future device-level works to fundamentally understand \proposal.}
\end{itemize}

\setcounter{version}{5}

\section{Background}
\label{sec:bkg}
We provide a concise overview of 1) DRAM organization and operation (\secref{sec:dram_organization_operation}), and 2) DRAM read disturbance (\secref{sec:dram_read_disturbance}).

\subsection{DRAM Organization and Operation}
\label{sec:dram_organization_operation}

\begin{figure}[h]
    \centering
    \includegraphics[width=\linewidth]{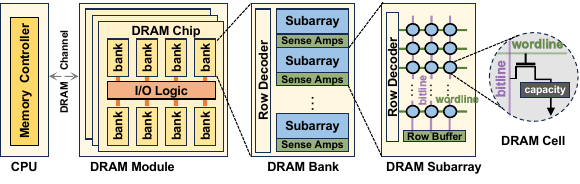}
    \caption{Hierarchical organization of modern DRAM.}
    \label{fig:dram_org}
\end{figure}

\noindent\textbf{Organization.} Fig.~\ref{fig:dram_org} \jwcr{1}{shows} the hierarchical organization of modern DRAM-based main memory. \jwcr{1}{The CPU's \emph{memory controller} communicates with \omcr{3}{one or more} \emph{DRAM module\omcr{4}{s}} over a \emph{memory channel}. A module contains one or \omcr{3}{more} \emph{DRAM ranks} that share the memory channel.} \jwcr{1}{Each} rank \omcr{3}{comprises} a set of \emph{DRAM chips} that \jwcr{1}{operate} in lockstep (i.e., all chips receive and process the same command \jwcr{1}{simultaneously}). Each DRAM chip \jwcr{1}{contains} multiple \emph{DRAM banks}, each consisting of many \emph{subarrays}\omcr{4}{~\cite{kim2012case,lee2013tldram,seshadri2013rowclone,chang2014improving}}. Inside a subarray, DRAM cells \jwcr{1}{are organized in a two-dimensional array connected by \emph{wordlines} and \emph{bitlines}}. A DRAM cell stores one bit of data \jwcr{1}{as electrical charge in a capacitor accessed through a transistor}. A row of cells \jwcr{1}{connected to} the same wordline is called a DRAM \emph{row}. A bitline connects the DRAM cells in the same column to a \emph{sense amplifier}. \jwcr{1}{The set of sense amplifiers storing} the data of an accessed row is \jwcr{1}{called} the \emph{row buffer}.

\noindent\textbf{Operation.} Accessing data in DRAM involves three main steps. First, the memory controller issues an activate command (\emph{ACT}) with a row address. The row decoder activates the corresponding row by driving its wordline, enabling the access transistors of the cells in that row. The data stored in the activated cells is sensed and transferred to the row buffer \jwcr{1}{through} the bitlines. Second, once the data are loaded into the row buffer, the memory controller issues \emph{RD}/\emph{WR} commands to read or write data \jwcr{1}{in the row buffer}. Third, before accessing \jwcr{1}{another row in} the same bank, the memory controller issues a precharge command (\emph{PRE}) to close the current row and restore the bitlines to the precharged state.

DRAM cells leak charge over time, risking \emph{retention\omcr{3}{-}failure} induced bitflips\omcr{3}{~\cite{khan2016parbor,liu2012raidr,liu2013retention,khan2014efficacy}} if \omcr{3}{the stored charge} is not restored promptly. To avoid this, the memory controller periodically \jwcr{1}{restores} each DRAM row's charge by issuing \emph{REF} (refresh) commands. Before issuing a REF command, the memory controller must send a \emph{PRE} command to close any open rows \jwcr{1}{to prepare the bank for refresh}.

\label{sec:bg:timing}
\noindent\textbf{Timing Parameters.} \jwcr{1}{To ensure reliable DRAM operation, the memory controller must obey standard timing parameters\omcr{3}{~\cite{lee2015aldram,kim2012case}}, which provide the DRAM circuitry sufficient time to complete the operations dictated by DRAM commands.} We describe~\rbm{five} timing parameters relevant to this work:~\rbm{(1)}~\emph{tRAS} is the minimum \jwcr{1}{interval} between an \emph{ACT} and a subsequent \emph{PRE} command. \rbm{(2)}~\emph{tRP} is the minimum interval between a \emph{PRE} and \rbm{a} following \emph{ACT} command. (3)~\emph{tRC} is the minimum interval between consecutive activat\omcr{3}{ing} commands to the same bank. 
(4)~\emph{tREFI} is the minimum interval between two consecutive \emph{REF} commands.~(5)~\emph{tREFW} is the maximum allowable interval between two refresh operations targeting the same row. Detailed explanations of these parameters
can be found in\omcr{3}{~\cite{jedec2012ddr3,jedec2020ddr4,jedec2024ddr5,lee2013tldram,lee2015aldram,kim2012case,chang2014improving}}.

\subsection{DRAM Read Disturbance}
\label{sec:dram_read_disturbance}

Read disturbance is \omcr{3}{the} phenomenon \omcr{3}{where} reading data from a memory or storage device causes physical disturbance (e.g., voltage deviation, electron injection, electron trapping) \jwcr{1}{on another piece of data that is \emph{not accessed} but located physically nearby the accessed data. Two prime examples of read disturbance in modern DRAM chips are} RowHammer\omcr{3}{~\cite{kim2014flipping,mutlu2017rowhammer,mutlu2019rowhammer, kim2020revisiting, frigo2020trrespass, orosa2021deeper, hassan2021utrr, kogler2022half, mutlu2023fundamentally, luo2024experimental, luo2025revisiting,olgun2025vrd,yaglikci2024svard}} and RowPress\omcr{3}{~\cite{luo2023rowpress,luo2024rowpress,luo2024experimental,luo2025revisiting,olgun2025vrd,yaglikci2024svard}}, \jwcr{1}{where repeatedly accessing (hammering) or keeping active (pressing) a row induces bitflips in physically nearby rows, respectively.} We refer to the unaccessed row \jwcr{1}{that experiences} bitflips as the \emph{victim row}, and the accessed row causing disturbances as the \emph{aggressor row}. 
\jwcr{1}{For read disturbance bitflips to occur, 1) the aggressor row needs to be activated more times than a certain threshold value (i.e., \threshold)~\cite{kim2014flipping}, and/or 2) the aggressor row needs to be open for a long period of time (i.e., \tagg{})~\cite{luo2023rowpress}.}
\vspace{0.25em}
\noindent\textbf{DRAM Read Disturbance Mitigation Mechanisms.} Many prior works propose mitigation techniques\omcr{4}{~\mitigatingRowHammerAllCitations} to protect DRAM chips against read disturbances. These methods typically involve: 1) detecting and recording the activation counts of potential aggressor rows and proactively refreshing victim rows before activation counts reach a critical threshold\omcr{4}{~\cite{kim2014flipping,qureshi2022hydra,bostanci2024comet,yaglikci2021blockhammer,kim2022mithril,saxena2022aqua,canpolat2025chronus,olgun2024abacus,seyedzadeh2016counter,park2020graphene,canpolat2024breakhammer,frigo2020trrespass,saxena2024start,qureshi2024mint,saxena2024impress}}, 2) selectively throttling accesses to aggressor rows to reduce their disturbance potential\omcr{3}{~\cite{yaglikci2021blockhammer,canpolat2024breakhammer,greenfield2016throttling}}.

\label{sec:bg:prac_rfm}
\vspace{0.25em}
\noindent\textbf{JEDEC Standard Mitigation.}~\omcr{4}{The} JEDEC DDR5 standard~\cite{jedec2020ddr5,jedec2024ddr5} introduces Refresh Management (RFM) and Per-Row Activation Counting (PRAC) to protect DRAM chips against read disturbance.
RFM~\cite{jedec2020ddr5} is a DRAM command that \jwcr{1}{grants} the DRAM chip a time window (e.g., 195 ns~\cite{jedec2020ddr5}) to proactively refresh potential victim rows.
The memory controller issues RFM commands, while the DRAM chip identifies and refreshes victim rows.
PRAC\omcr{3}{~\cite{jedec2024ddr5,canpolat2025chronus}} implements an activation counter for each DRAM row and thus accurately
tracks the activation counts of all rows~\cite{jedec2024ddr5}. When a row’s activation count reaches a configurable fraction of threshold (i.e., alert back-off threshold, \emph{ABO threshold}, where the fraction can be configured to
either 70\%, 80\%, 90\%, or 100\%\cite{jedec2024ddr5}), the DRAM chip asserts a back-off signal (\emph{ABO}) which forces the memory controller to issue an RFM command. The DRAM chip proactively refreshes potential victim rows upon receiving an RFM command.
\setcounter{version}{3}
\section{\proposal Characterization}
\jwcr{1}{We describe our commercial off-the-shelf (COTS) DRAM testing infrastructure (\secref{sec:infra}) and testing methodology (\secref{sec:method}).}

\subsection{DRAM Testing Infrastructure}
\label{sec:infra}

\begin{figure}[h]
    \centering
    \includegraphics[width=\linewidth]{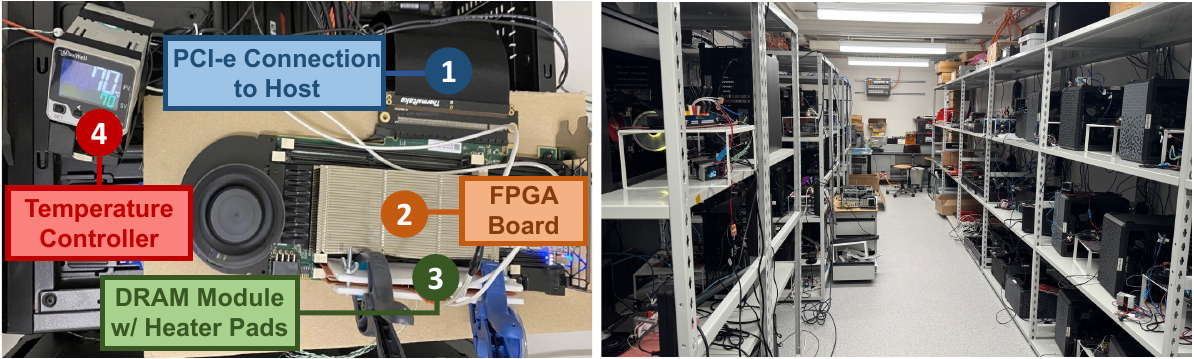}
    \caption{\omcr{2}{FPGA-based DRAM and memory controller testing setup (left) and laboratory (right).}}
    \label{fig:fpga}
\end{figure}

We \jwcr{1}{conduct COTS DRAM chip experiments using DRAMBender~\cite{olgun2023drambender,safari-drambender} (built \omcr{3}{on} SoftMC~\cite{hassan2017softmc,softmc_github}), an FPGA-based testing infrastructure that provides precise control of DDR4 and HBM2 commands issued to DRAM modules. \figref{fig:fpga} \omcr{2}{(left)} shows our DDR4 experimental setup}, which comprises four main components: 1) a host machine that generates the test program and collects experimental results, 2) an FPGA board~\cite{xilinx_alveo_u200}, 3) a thermocouple temperature sensor and a pair of heater pads mounted on the DRAM chips to {maintain a target temperature level}, and 4) a temperature controller~\cite{maxwell_ft20x} to control the temperature with a precision of ± 0.5$^{\circ}$C. \omcr{2}{\figref{fig:fpga} (right) shows our laboratory comprising many DDR4/HBM2 testing platforms.}

\noindent\textbf{Real DDR4 and HBM2 DRAM Chips Tested.} Table~\ref{tab:chips} \jwcr{1}{provides 196 real DDR4 chips from 20 modules and 3 HBM2 chips that we characterize}.

\vspace{0.5em}
\begin{table}[ht]
\caption{\small\textbf{Summary of DDR4 and HBM2 DRAM chips tested.}}
\label{tab:chips}
\resizebox{\linewidth}{!}{%
\renewcommand{\arraystretch}{0.92}
\begin{threeparttable}
\begin{tabular}{cccccc}
\textbf{Chip Mfr.} & \textbf{Module IDs} & \textbf{\#Chips} &  \textbf{Die Rev.} & \textbf{Density} & \textbf{Org.}  \\
\hline
\toprule
                        & S0, S1 & 8 & B & 16Gb & x8 \\
Mfr. S                  & S2, S3, S4 & 8 & A & 16Gb & x8 \\
(Samsung)               & S5, S6 & 8 & D & 8Gb & x8  \\
                        & S7 & 16 & C & 8Gb & x4 \\
                        & S8 & 8 & F & 4Gb & x8 \\ 
\midrule            
                        & H0 & 16 & A & 16Gb & x8 \\
Mfr. H                  & H1 & 8 & D & 8Gb & x8  \\
(SK Hynix)              & H2 & 8 & C & 8Gb & x8  \\
                        & H3 & 16 & C & 8Gb & x8  \\
                        & H4 & 8 & C & 4Gb & x8 \\

\midrule
                        & M0 & 16 & F & 16Gb & x8 \\
Mfr. M                  & M1 & 8 & F & 16Gb & x8  \\
(Micron)                & M2, M3 & 4 & E & 16Gb & x16  \\
                        & M4 & 4 & B & 16Gb & x16  \\
                        & M5 & 16 & E & 8Gb & x8  \\ 
\midrule
Samsung     & HBM2 Chips  & 3 & N/A  & N/A & N/A \\
   
\bottomrule
\hline
\end{tabular}
\begin{tablenotes}
\item[a] We report “N/A” if the information is not publicly available.
\end{tablenotes}
\end{threeparttable}
}
\end{table}  
\vspace{-0.5em}

\subsection{Experimental Methodology}
\label{sec:method}

\noindent\textbf{Disabling Sources of Interference.} We identify three factors that can affect our results: 1) data retention failures~\cite{liu2013retention,patel2017reaper}, 2)~on-DRAM-die read disturbance mitigation mechanisms (e.g., Target Row Refresh (TRR)~\cite{frigo2020trrespass,hassan2021utrr,micron2018ddr4}), and 3) error correction codes (ECC)\omcr{2}{~\cite{patel2020beer,patel2021harp,patel2019understanding,kang2014co}}. We \jwcr{1}{follow} prior work~\cite{olgun2025vrd,luo2024experimental,olgun2024hbm,orosa2021deeper,yaglikci2024svard} to eliminate \jwcr{1}{these} interference: i) we \jwcr{1}{}{complete} all experiments within a single refresh window to avoid \jwcr{1}{retention-failure} bitflips~\cite{jedec2020ddr4,jedec2012ddr3}, ii) we disable periodic refresh\jwcr{1}{es} to disable TRR \jwcr{1}{mechanisms}, \rbm{and} iii) we \jwcr{1}{ensure all tested DRAM} modules have neither rank-level nor on-die ECC. These steps ensure that we directly observe all circuit-level bitflips without interference from architectur\jwcr{1}{al} \jwcr{1}{mitigation or correction} techniques.

\noindent\textbf{Logical-to-Physical Row Mapping.} DRAM manufacturers \rbm{internally map logical addresses to} physical addresses\omcr{2}{~\cite{kim2012case,lee2013tldram,seshadri2013rowclone,chang2014improving,kim2014flipping,barenghi2018rowmap,tatar2018defeating,cojocar2020arewe}}. To identify victim rows and their physically adjacent aggressor rows, we reverse engineer the row mapping scheme using \jwcr{1}{the} methodology in~\cite{kim2020revisiting,orosa2021deeper,barenghi2018rowmap,yaglikci2022understanding,luo2023rowpress,olgun2025vrd,yaglikci2024svard,luo2024experimental,tugrul2025understanding}.

\noindent\textbf{Coupled Rows.}
\omcr{2}{Prior works~\cite{nam2023xray,nam2024dramscope} show that DDR4 chips may contain coupled rows, where two distinct DRAM rows in the same bank can be simultaneously activated with a single DRAM command. 
This unintended row activation can disturb bitflips in the victim row\omcr{2}{~\cite{baek2025marionette}}.
To avoid this effect, we identify non-coupled rows using single-sided RowHammer~\cite{kim2014flipping,kim2020revisiting,seshadri2013rowclone,mutlu2019rowhammer} and restrict our experiments to non-coupled rows only.}

\noindent\textbf{True and Anti \jwarv{2}{C}ells.} True~\rbm{and anti} cells {determine} whether a charged capacitor \rbm{represents a} logical 1 or \rbm{a} logical 0. \jwcr{1}{Prior} works~\omcr{2}{\cite{liu2013retention,nam2024dramscope,nam2023xray,marazzi2024hifi,wu2019protecting}} on DRAM retention failure commonly assume that DRAM retention-induced errors are only 1 to 0 bitflips. We use this finding to reverse engineer the true/anti layout of each module in our experiment and align the expected data patterns with the actual charge polarity.

\begin{figure}[h]
    \centering
    \includegraphics[width=\linewidth]{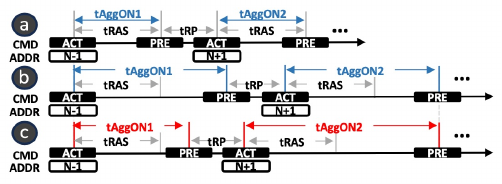}
    \caption{\jwcr{1}{Three DRAM access patterns:} (a) double-sided RowHammer, (b) double-sided RowPress, and (c) \proposal.}
    \label{fig:access_pattern_comparison}
\end{figure}

\noindent\textbf{Access Pattern.} Fig.~\ref{fig:access_pattern_comparison} \jwcr{1}{shows} three distinct memory access patterns: (a) double-sided RowHammer~\cite{kim2014flipping}, (b) double-sided RowPress~\cite{luo2023rowpress}, and (c) the proposed \proposal. In each case, two aggressor rows (\rowagx~and \rowagy) \rbm{target a victim row (\rowv)} to induce bitflips. In each access loop \omcr{2}{(i.e., one complete access iteration over both aggressor rows)}, we (i) activate the aggressor row (\rowagx) \rbm{for a duration of} \taggone{}, (ii) \rbm{precharge the bank} and wait for \tRP{}, (iii) activate~\rbm{the} aggressor row (\rowagy) \rbm{for a duration of} \jwcr{1}{\taggtwo{}}, and (iv) precharge \rbm{the bank again} and wait \rbm{for} \tRP{}. Notably,\revci{A1, A2} in double-sided RowHammer, both \jwcr{1}{\taggone{}} and \jwcr{1}{\taggtwo{}} are set to {\jwcr{1}{\tRAS}}. In double-sided RowPress, both aggressor rows remain open for an equally prolonged duration (i.e., \jwcr{1}{\taggone{}} = \jwcr{1}{\taggtwo{}} $>$ \tRAS{}). In contrast, \proposal maintains the same total row \jwcr{1}{open} time per access loop as double-sided RowPress but asymmetrically distributes row open times between the two aggressor rows (i.e., \jwcr{1}{\taggone{}} is not necessarily \jwcr{1}{equal} to \jwcr{1}{\taggtwo{}}).

\noindent\rbm{\textbf{Open Time Budget.} To clarify the row open time distribution {in} \proposal access pattern, we define \emph{Open Time Budget (OTB)} as:}

\vspace{-1em}
\begin{equation}
\begin{aligned}
    \mathbf{OTB} = & \mathbf{t_\textit{AggON1}} + \mathbf{t_\textit{AggON2}} - 2 \cdot \mathbf{t_{\textit{RAS}}}
    &\textcolor{black}{\text{s.t. } \mathbf{t_\textit{AggON}} \ge \mathbf{t_\textit{RAS}}}
\end{aligned}
\end{equation}

\begin{figure}[h]
    \centering    
    \includegraphics[width=1.0\linewidth]{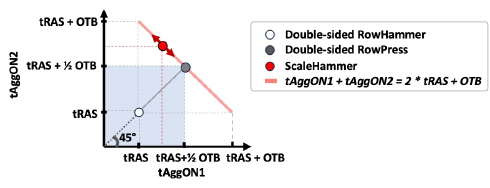}
    \caption{\jwcr{1}{\proposal Design Space}}
    \label{fig:otb_assign}
\end{figure}

\noindent By design,~\rbm{both \jwcr{1}{\taggone{}} and \jwcr{1}{\taggtwo{}} are greater than or equal to \jwcr{1}{\tRAS{}}}, ensuring that OTB \rbm{accounts for} the additional open time budget.\footnote{\rbm{When} \jwcr{1}{\taggone{}} = \jwcr{1}{\taggtwo{}},~\proposal is identical to double-sided \rbm{RowPress. When} OTB = 0, \proposal is identical to double-sided RowHammer.} \figref{fig:otb_assign} shows the \proposal design space for allocating a shared row open-time budget between two aggressor rows. While double-sided RowHammer (white circle) and double-sided RowPress (grey circle) are restricted to the $45^\circ$ symmetry line (grey line), \proposal (red circle with double sided arrow) explores the asymmetric configurations along the constant OTB boundary (slope = -1).
We use six OTB\revci{E5.1} values to span a wide range of activation durations: in DDR4, 48~ns is close to \jwcr{1}{\tRC{}}, representing near-nominal activation timing, while 7.8~$\mu s$ corresponds to \textit{tREFI}, capturing an open time bound aligned with the refresh interval. The intermediate values (i.e., 120, 300, 600, \omcr{2}{\qty{1200}{ns}}) increase approximately logarithmically to capture the transition from short to long.

\noindent\textbf{Metrics.} \rbm{To \jwcr{1}{quantify} how \proposal amplifies DRAM's vulnerability to read disturbance, we examine how \threshold changes as \jwcr{1}{\taggone{}} increases \jwcr{1}{under} a \jwcr{1}{fixed} OTB.} A lower \threshold indicates that a victim row {exhibits greater vulnerability}. 
For each OTB, \jwcr{1}{we sweep \taggone{} from \tRAS{} to \tRAS{} + OTB} in steps of 5\% of the OTB, and \jwcr{1}{report the minimum observed \threshold as the \emph{final} \threshold{}}. For each module, we randomly select 512 rows in bank 1 and independently characterize their vulnerability under different access patterns.

\noindent\textbf{Algorithm.} We \jwcr{1}{measure a row's} \threshold using a bisection \omcr{2}{search} method used by prior works~\omcr{2}{\cite{kim2014flipping,yaglikci2022understanding,orosa2021deeper,kim2020revisiting,luo2023rowpress,luo2024rowpress,luo2024experimental,olgun2024hbm,tugrul2025understanding}}. 
\omcr{2}{We use a fixed \threshold accuracy of 300 (i.e., we terminate the search for \threshold when the difference between the current and previous measurements of \threshold is no larger than 300).}
We repeat \jwcr{1}{our test}~\emph{three times} and report the minimum \threshold.

\noindent\textbf{Data Pattern.} We use two data patterns%
,\footnote{Using all-zero or all-one data patterns eliminates dependence on the complicated column mapping, which manufacturers do not publicly disclose, and ensures that the physical row and its cells hold the expected values.} \textit{Rowstripe0} (victim row\jwcr{1}{s store all zeros and aggressor rows store all ones}) and \textit{Rowstripe1} (victim row\jwcr{1}{s store all ones and aggressor rows store all zeros}), \rbm{that are} widely used in memory reliability testing and prior DRAM characterization work\jwarv{2}{s}~\cite{van2002address, kim2014flipping,luo2023rowpress,olgun2024hbm,yaglikci2024svard,kim2020revisiting,orosa2021deeper,yaglikci2022understanding,tugrul2025understanding,luo2024rowpress,luo2025revisiting,luo2024experimental,lee2021cryoguard}.

\noindent\textbf{Temperature.} We \jwcr{1}{conduct our experiments at} \qty{50}{\celsius} and \qty{80}{\celsius}.

\setcounter{version}{3}
\section{Characterization Results}
\jwcr{1}{We demonstrate real DDR4 and HBM2 DRAM chips' vulnerability to read disturbance under \omcr{2}{the \proposal access pattern}. \proposal{} amplifies DRAM\omcr{2}{'s} vulnerability to read disturbance, widely affects rows in DRAM chips from three major DRAM manufacturers, and worsens with DRAM technology scaling \omcr{2}{(}\secref{sec:ddr4_char}\omcr{2}{)}. We \omcr{3}{study} \proposal{}'s characteristics by distinguishing it from Variable Read Disturbance~\cite{olgun2025vrd}, analyzing \omcr{3}{bitflip} locations, comparing it with other access patterns \omcr{2}{(}\secref{sec:distinguish}\omcr{2}{)}, and \omcr{2}{characterize} \proposal{} on HBM2 DRAM chips \omcr{2}{(}\secref{sec:hbm2}\omcr{2}{)}.}

\subsection{DDR4 Chips' Vulnerability to Read Disturbance}
\label{sec:ddr4_char}

\iey{We investigate the variation in \threshold{} across rows when we perform \proposal and double-sided RowPress. \figref{fig:acmin_reduction} (left) shows the distribution of \threshold under \proposal, \jwcr{1}{normalized} to double-sided RowPress \jwcr{1}{(y-axis), across} six OTB values (hue).} The x-axis shows the percentile of victim rows, ranked in ascending order of \threshold reduction. \omcr{2}{V}ictim rows toward the right side of the plot experience larger \threshold reductions.
For example, at OTB = 7800 ns (blue line), a normalized \threshold of 0.6 at P75 indicates that 75\% of victim rows have normalized \threshold{} \jwcr{1}{values} above 0.6, while the remaining 25\% fall below 0.6.  
\figref{fig:acmin_reduction} (right) shows \threshold (y-axis) \jwcr{1}{under} \proposal and double-sided RowPress across OTB values (x-axis). We make two key observations:

\begin{figure}[h]
    \centering
    \includegraphics[width=\linewidth]{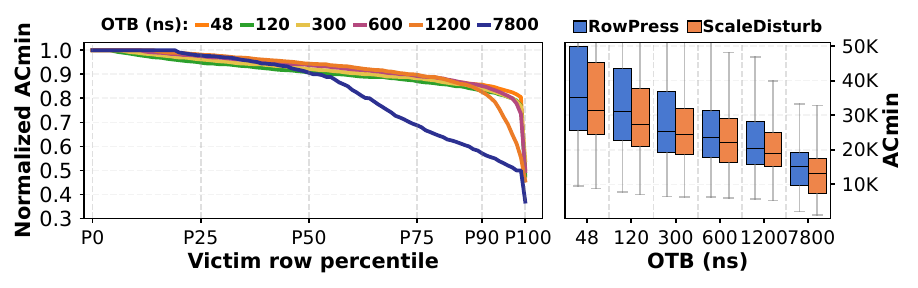}
    \caption{Normalized \threshold{} distribution across \jwcr{1}{all} victim rows, sorted by reduction percentile, for six OTB values (left) and \threshold{} \jwcr{1}{distribution under} RowPress and \proposal (right)}
    \label{fig:acmin_reduction}
\end{figure}

\begin{obsvbox}
{\proposal reduces \threshold by \meanavr{} on average (up to \maxavr) \omcr{2}{across all tested rows} compared to double-sided RowPress.}
\end{obsvbox}

\jwcr{1}{Across all \omcr{2}{tested} rows under six OTB values,} the normalized \jwcr{1}{\threshold decreases} from 1.0 at P0, to \jwcr{1}{0.904} at P50, and drops \jwcr{1}{sharply} beyond P90, reaching as low as 0.37 (\maxavr reduction at OTB = 7.8\,µs). \jwcr{1}{As} \proposal and double-sided RowPress use the same \omcr{2}{OTB}, reduction\jwcr{1}{s} in \threshold proportionally \jwcr{1}{reduce} the read disturbance time required to induce \omcr{2}{read disturbance} bitflips.

\begin{obsvbox}
\proposal reduces the minimum \threshold at the module level by \meanmvr{} \jwcr{1}{on average} (up to 52.4\%) compared to double-sided RowPress.
\end{obsvbox}

We analyze the module-level minimum \threshold (i.e., the lowest \threshold across all rows in a module), \jwcr{1}{because it determines the threshold required for read disturbance mitigation}. \proposal reduces this value by 16.1\% on average under relatively small OTBs (e.g., $<$\,\qty{1200}{ns}), and by up to 52.4\% under large OTB (7.8\,$\mu$s), compared to double-sided RowPress.

\begin{figure}[h]
    \centering
    \includegraphics[width=\linewidth]{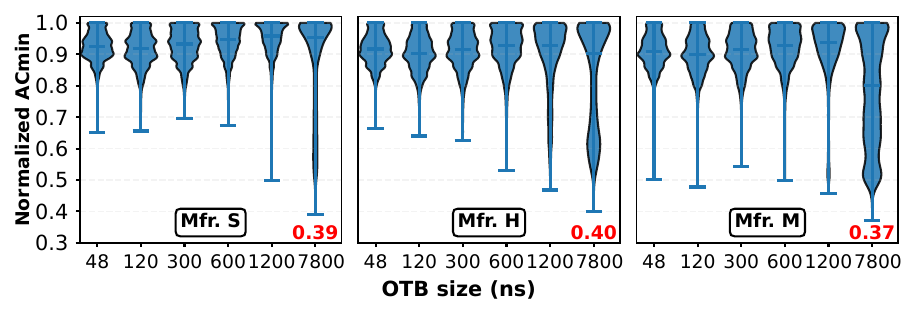}
    \caption{Normalized \threshold distributions across different OTB sizes for DRAM chips from manufacturers: Samsung, SK Hynix and Micron}
    \label{fig:reduction_mfr}
\end{figure}

Fig.~\ref{fig:reduction_mfr} shows the normalized~\threshold (y-axis) distribution across \rbm{chips} from Samsung, SK Hynix, and Micron. Each violin \omcr{2}{plot} represents \omcr{2}{the distribution of} \threshold{} \jwcr{1}{at} a given OTB (x-axis)\omcr{2}{: whiskers indicate the minimum and maximum values, the dashed line inside denotes the median, and the width at each point reflects the distribution density.}
We \jwcr{1}{highlight the minimum normalized \threshold for each manufacturer} in red \omcr{3}{text}.

\begin{obsvbox}
\proposal widely affects DRAM chips from all three major DRAM manufacturers.
\end{obsvbox}

\proposal{} reduces \threshold by 7.8\%, 10.4\%, and 11.2\% on average, and by up to 61\%, 60\%, and 63\% for DRAM chips from Samsung, SK Hynix, and Micron, respectively.
Notably, \omcr{2}{when OTB $<$\,\qty{120}{ns},} \proposal{} \omcr{2}{reduces} \threshold by up to \jwcr{1}{35}\%, \jwcr{1}{33}\%, and 50\% \jwcr{1}{for chips from Samsung, SK Hynix, and Micron, respectively}.
In this range, \jwcr{1}{RowPress} disturbance \omcr{2}{is} limited, \omcr{2}{as prior} device-level studies~\cite{zhou2024unveiling,zhou2024understanding} \omcr{2}{show that the impact of the underlying electric field weakens with shorter row open times}. \jwcr{1}{In contrast}, \proposal asymmetrically allocates the row open times, disrupting field balance and amplifying disturbance even under the same OTB (\secref{sec:physical_explanation}). \omcr{2}{\proposal's} \jwcr{1}{effectiveness at short OTBs} makes \jwcr{1}{malicious} attacks more practical in real systems, where rows are \omcr{2}{typically not kept} open for long due to access conflicts and memory controller scheduling.

\begin{figure}[h]
    \centering
    \includegraphics[width=\linewidth]{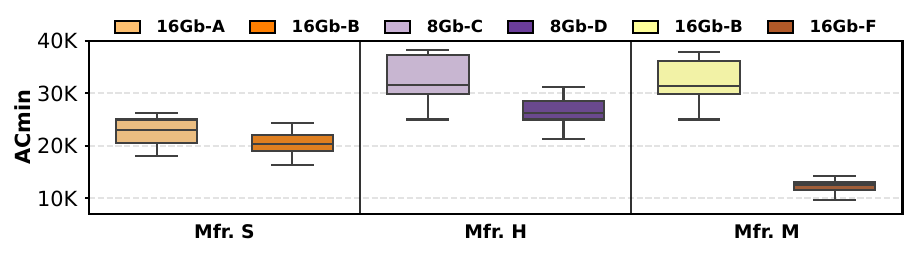}
    \caption{\threshold distributions across different density-revision chips}
    \label{fig:node_scaling}
\end{figure}

We investigate the impact of DRAM technology scaling on \proposal. 
Fig.~\ref{fig:node_scaling} shows \threshold distribution (y-axis) across chips with different densities and die revisions (hue) from Samsung, SK Hynix, and Micron, at a practical \qty{120}{ns} OTB. \omcr{2}{Each box plot shows the distribution of \threshold{} across all tested rows for each die revision: whiskers denote the minimum and maximum values, and the dashed line indicates the median.}

\begin{obsvbox}
Newer chips tend to be more vulnerable to \proposal bitflips.
\end{obsvbox}

We observe that in the same die density for each tested manufacturer, more advanced technology nodes\footnote{The technology node used in manufacturing a DRAM chip is not publicly available information. Prior works~\omcr{2}{\cite{kim2014flipping,kim2020revisiting,luo2023rowpress,orosa2021deeper,mutlu2017rowhammer,mutlu2019rowhammer,yaglikci2024svard, luo2024rowpress,tugrul2025understanding,yaglikci2022understanding,mutlu2023fundamentally}} assume that for a given chip manufacturer and die density, the alphabetical order of die revision codes may indicate technology node advancement.} experience higher \proposal vulnerability. The average \threshold under \proposal decreases by 9.8\%, 18.4\%, and 62.0\% for Samsung (16Gb, A to B), SK Hynix (8Gb, C to D), and Micron (16Gb, B to F), respectively. 
We hypothesize that reduced physical distances in more advanced chips enhance electric-field coupling and cell sensitivity to interference \omcr{2}{(see }\secref{sec:physical_explanation}\omcr{2}{)}. 

\begin{takeawaybox} 
\label{tkw:common}\proposal{} \rbm{(1) amplif\omcr{2}{ies}} DRAM's vulnerability to read disturbance, (2) widely affects DRAM chips, and (3) \rbm{becomes more severe with continued DRAM scaling}.
\end{takeawaybox}

\begin{figure}[!ht]
    \centering
    \includegraphics[width=\linewidth]{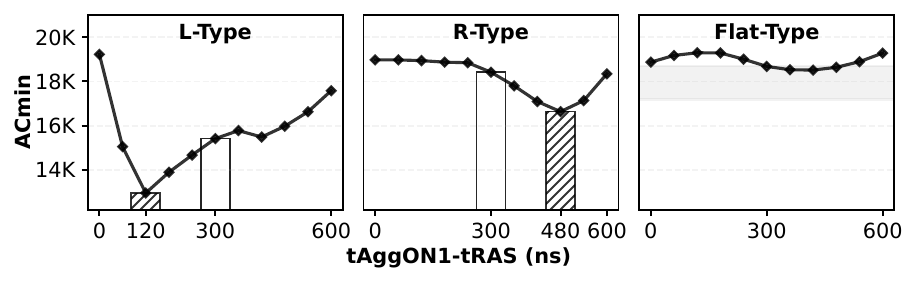}
    \caption{Three observed \threshold \omcr{2}{reduction} patterns: L-type, R-type and Flat-type, when sweeping \jwcr{2}{\taggone{}}.}
    \label{fig:three_patterns}
\end{figure}

\jwcr{1}{We further analyze the behavior of row-level \threshold as row open time varies under a fixed OTB.}
Fig.~\ref{fig:three_patterns} shows three types of trend\omcr{2}{s we observe:}\footnote{The three patterns are derived from three representative victim rows and describe all tested rows. Although the \threshold curves are somewhat irregular (due to coarse-grained sweeping), their overall trends are consistent. Statistically, Flat-Type rows show higher average \threshold than L- and R-Type rows.}
\omcr{3}{w}e use the \omcr{2}{middle bar (where \taggone{} = \taggtwo{})} as the reference and the lowest observed \threshold as the minimum (hatched bar). 
Rows with less than 10\% \threshold reduction \omcr{2}{under \proposal} are Flat-type; others are L- or R-type, based on the minimum \omcr{2}{\threshold}’s position relative to the \omcr{2}{middle bar}.

\begin{obsvbox}
DRAM rows under \proposal exhibit three \omcr{2}{major} \threshold \omcr{2}{reduction} patterns: L-, R- and Flat-type.
\end{obsvbox}

\omcr{2}{We find that as \taggone{} increases, the minimum \threshold{} of a row may occur at either lower (e.g., \qty{120}{ns}) or higher (e.g., \qty{480}{ns}) \taggone{} values, while some rows exhibit little change in \threshold{}.}
We hypothesize that \omcr{2}{vulnerable cells within a DRAM row experience unequal read disturbance from the two aggressor rows. Varying the row open time changes the accumulated disturbance (see~\secref{sec:physical_explanation}), leading to lower \threshold and different \threshold{} reduction patterns.}

\begin{figure}[!h]
    \centering
    \includegraphics[width=\linewidth]{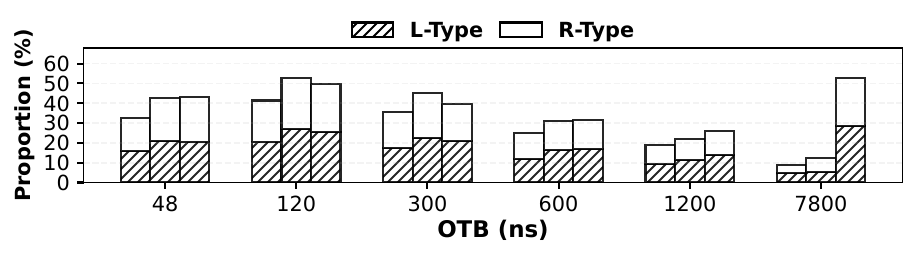}
    \caption{Proportion of \threshold \omcr{2}{reduction} pattern (L, R) for chips from Samsung, SK Hynix and Micron (three bars, left to right) across OTBs}
    \label{fig:pattern_proportion}
\end{figure}

Fig.~\ref{fig:pattern_proportion} shows the proportion of two \threshold{} \omcr{2}{reduction} patterns (L-Type and R-Type) across OTBs, \omcr{2}{with the remainder corresponding to the Flat pattern}. For each OTB, the three bars correspond to Samsung, SK Hynix, and Micron, respectively. \omcr{3}{We make two key observations:}

\begin{obsvbox}
L-type and R-type patterns constitute a large \omcr{2}{fraction} of rows, \omcr{2}{and} their proportions \omcr{2}{vary with} OTB.
\end{obsvbox}

L- and R-type patterns collectively account for an average of \rbi{37.5\% (up to 49.2\%)} of rows \rbm{across} all OTB values, indicating that a substantial fraction of rows are vulnerable to \proposal with over 10\% \threshold reduction.
The combined proportion increases with OTB, peaking at 49.2\% near \qty{120}{ns}, then declines beyond \qty{300}{ns} to below 10\% at \qty{7.8}{\micro\second}. Micron modules deviate from this trend, showing a\omcr{2}{n increase} after \qty{1200}{ns}.

\begin{obsvbox}
L-type and R-type patterns occur at nearly equal rates across all evaluated OTB values.
\end{obsvbox}

L-type and R-type patterns occur in nearly equal proportions, averaging 19.6\% and 17.9\%, respectively. 
\omcr{2}{We hypothesize that in} a saddle-fin\omcr{2}{-based} recessed channel structure~\cite{zhou2024unveiling,zhou2024understanding}, where \omcr{2}{the active} wordlines (\emph{AWL}) share the active region (AR) and \omcr{2}{the passing} wordlines (\emph{PWL}) are isolated \omcr{2}{by AR}, AWL–PWL or PWL–AWL alignment yields L-\omcr{2}{type} or R-type patterns, respectively.

\begin{takeawaybox}
\proposal alters \threshold via three OTB-influenced patterns. L-Type and R-type patterns consistently lower \threshold by over 10\%, with similar frequency.
\end{takeawaybox}

\subsection{Distinguishing Characteristics of \proposal}
\label{sec:distinguish}
\begin{figure}[!ht]
    \centering 
    \includegraphics[width=\linewidth]{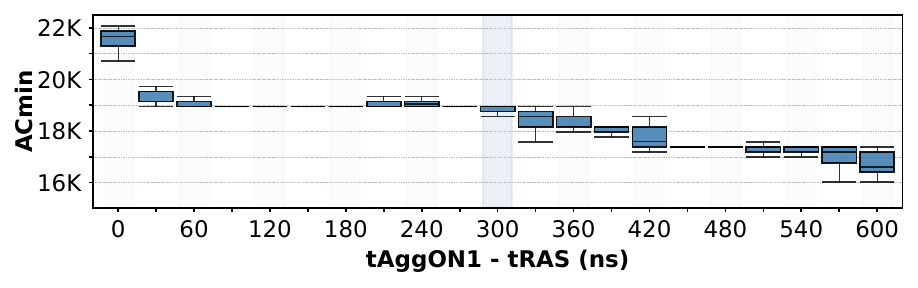}
    \caption{\rbm{\proposal stability: \threshold measured over 300 repeated tests for each \jwcr{2}{\taggone{}}, with OTB fixed at \qty{600}{ns}, on a representative victim row exhibiting the R-Type pattern.}}
    \label{fig:scale_vrd}
\end{figure}

\noindent\textbf{\jwcr{1}{Variable Read Disturbance (VRD)}}\omcr{3}{~\cite{olgun2025vrd,olgun2026discord}} is a recently identified phenomenon, \rbm{where the read disturbance threshold (RDT) of a row} \jwcr{1}{unpredictably changes} over time. 
To distinguish the observed \threshold{} reduction from VRD, we examine
\omcr{3}{1) the magnitude of the \threshold{} reduction induced by \proposal{},
and 2) the stability of the \threshold reduction across repeated tests.}
Fig.~\ref{fig:scale_vrd} \jwcr{1}{shows the \threshold distribution (y-axis) across 300 tests (\omcr{3}{each box plot}) at varying} \jwcr{2}{\taggone{}} (x-axis), with OTB fixed at \qty{600}{ns}, \jwcr{1}{for} a \omcr{2}{representative} victim row exhibiting R-Type pattern. The top and bottom whiskers denote the maximum and minimum \threshold. We alternate the background color for \jwcr{2}{\taggone{}} values. \omcr{3}{We make two key observations:}

\begin{obsvbox}
\omcr{2}{\proposal reduces \threshold more than \omcr{3}{Variable Read Disturbance}.}
\end{obsvbox}

Compared to the \threshold{} \omcr{3}{variation} induced by VRD within each box \omcr{3}{plot (i.e., \threshold{} differences across hundreds of measurements)}, \proposal reduces \omcr{3}{the average} \threshold{} \omcr{3}{by varying row open times:} from \omcr{3}{21678} at \taggone{} = \omcr{3}{\qty{0}{ns}}, to \omcr{3}{16598} at \taggone{} = \omcr{3}{\qty{600}{ns}}, \omcr{3}{representing a 23.4\% reduction in \threshold{}}.
In terms of \jwcr{1}{the} physical mechanism, VRD attributes the \omcr{3}{temporal variation} in \threshold to the randomly changing occupancy of charge traps in the shared active region of the aggressor and victim cells~\cite{olgun2025vrd}. \omcr{3}{W}e hypothesize that the reduction \omcr{2}{in \threshold due to \proposal} is driven by electric-field interference, \jwcr{1}{where \proposal increases the net \omcr{2}{electric} field by keeping rows open for asymmetric durations (see \secref{sec:physical_explanation})}.

\begin{obsvbox}
{\threshold reduction caused by \proposal is stable across our hundreds \rbm{of} tests.}
\end{obsvbox}

\omcr{3}{Across our 300 repeated tests, the \threshold{} reduction across different \taggone{} values remains stable. For example, the average \threshold{} values measured across all 300 tests are 21678 at \taggone{} = \qty{0}{ns}, 18944 at \taggone{} = \qty{300}{ns}, and 16598 at \taggone{} = \qty{600}{ns}, demonstrating consistent \threshold{} differences among different \taggone{} configurations.}

\begin{takeawaybox}
\proposal is fundamentally different from \omcr{2}{Variable Read Disturbance} in both stability and \omcr{2}{\threshold reduction}.
\end{takeawaybox}

\begin{figure}[h]
    \centering\includegraphics[width=\linewidth]{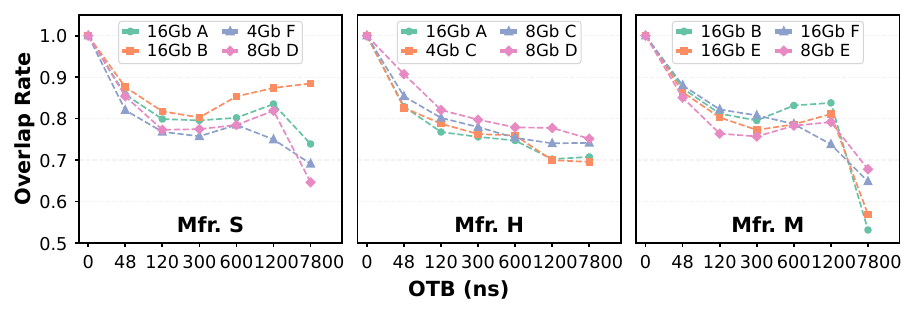}
    \caption{Overlap of first-bitflip \omcr{2}{locations}.}
    \label{fig:overlap}
\end{figure}

\noindent\textbf{Bitflip Index Analysis.} 
We investigate the indices of the first flipped cell(s) \jwcr{1}{induced by \proposal}. Fig.~\ref{fig:overlap} shows the overlap rate (y-axis), defined as the fraction of rows \omcr{2}{whose first flipped cell is the same} under both \proposal{} and double-sided RowPress, across OTBs (x-axis) and manufacturers (subplots), with modules grouped by die revision (hue). \jwcr{1}{We make two key observations:}

\begin{obsvbox}
{\proposal} flips a distinct set of cells \omcr{2}{compared to} double-sided RowPress.   
\end{obsvbox}

\jwcr{1}{The \omcr{2}{first flipped cell} within a row differ\omcr{2}{s} significantly between \proposal and double-sided RowPress across a large fraction of rows.} For example, the overlap ratio of 16Gb B-Die \jwcr{1}{chips from} Micron \jwcr{1}{decreases} to 0.52 at \qty{7.8}{\omcr{2}{\micro}\second}, {indicating} that 48\% of rows flip at entirely different cells. 

\begin{obsvbox}
The difference in the \omcr{3}{locations of the first} flipped cells between \proposal and RowPress \omcr{3}{increases} with OTB.
\end{obsvbox}

\jwcr{1}{\omcr{3}{T}he average overlap rate decreases from 1 to 0.78, 0.77, and 0.79 at OTB = \omcr{2}{\qty{300}{ns}}, and further to 0.72, 0.72, and 0.61 at \qty{7.8}{\omcr{2}{\micro}\second}, for chips from Samsung, SK Hynix, and Micron, respectively.}
\jwcr{1}{Our analysis of bitflip indices across all tested rows shows that 1) each victim row exhibits a \omcr{2}{fixed} set of vulnerable cells (one or more) for a given OTB under \proposal, and 2) the first flipped cell in a victim row may \omcr{2}{change} within this set as \jwcr{1}{\taggone{}} changes.}
We attribute the \omcr{2}{change} of the first flipped cell \jwcr{1}{within the victim row} (under a fixed OTB) to \jwcr{1}{variations in cell vulnerability} to asymmetric electric fields.

\begin{takeawaybox}
\proposal and double-sided RowPress flip distinct cell\omcr{2}{s} and their difference increases with OTB values.
\end{takeawaybox}

\noindent\textbf{Access Pattern Analysis.}~We evaluate \omcr{3}{DRAM rows'} vulnerability~\omcr{3}{to read disturbance} under different access patterns: double-sided RowPress\omcr{3}{~\cite{luo2023rowpress,luo2024rowpress,luo2024experimental}}, single-sided RowPress (SSRP)\omcr{3}{~\cite{luo2023rowpress,luo2024rowpress,luo2024experimental}}, \jwcr{1}{double- and single-sided RowHammer}\omcr{3}{~\cite{kim2014flipping,kim2020revisiting}}, and \proposal. Fig.~\ref{fig:ssrp} shows \omcr{3}{the} \threshold{} distribution (y-axis) across varying OTB values (x-axis) under different access patterns (hue), \jwcr{1}{and each subplot corresponds to a representative module from a different manufacturer}.
We label OTB = 0 as "RowHammer"~\omcr{3}{(highlighted with a yellow background)} \jwcr{1}{and evaluate} single-sided \jwcr{1}{patterns} separately for the upper and lower aggressor rows. 
We make two key observations:

\begin{figure}[!ht]
    \centering
    \includegraphics[width=1.0\linewidth]{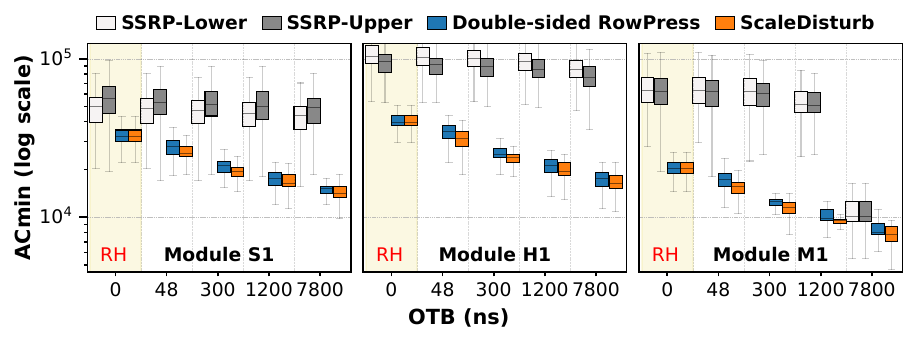}
    \caption{\threshold distribution under different access patterns.}
    \label{fig:ssrp}
\end{figure}

\begin{obsvbox}
Victim rows are asymmetrically vulnerable to upper and lower aggressors under single-sided patterns.
\end{obsvbox}
\omcr{2}{For} all \omcr{2}{tested} OTBs, we observe average \threshold variations of 14.0\%, 11.2\% and 3.0\%
\omcr{2}{between} applying single-sided patterns to the upper \omcr{2}{and} lower aggressor rows, for modules S1, H1, and M1, respectively.
This indicates that the effects of two adjacent aggressor rows on the victim row are not strictly symmetric. We attribute it to the uneven wordline layout (i.e., AWL and PWL), which might lead to differences in charge paths and electric field strength, thereby 
\omcr{2}{causing differences in charge displacement in the victim cell}.

\begin{obsvbox}
\proposal consistently lowers~\threshold{} \omcr{2}{compared to} single- and double-sided RowHammer/RowPress~\omcr{3}{access patterns.}
\end{obsvbox}

\jwcr{1}{\omcr{2}{For} all tested OTBs, double-sided patterns are more effective than single-sided patterns,} and prolonged \tagg{} amplifies disturbance. 
\jwcr{1}{\proposal reduces \threshold by 58.0\%, 74.0\%, and 67.3\% on average compared to single-sided RowPress, and by 6.0\%, 6.7\%, and 9.3\% compared to double-sided RowPress, for modules
S1, H1, and M1, respectively.}

\begin{takeawaybox}
Single-sided access {patterns} causes asymmetric vulnerability based on {victim row's} position. \proposal consistently \omcr{2}{leads to} the lowest \threshold across \omcr{2}{all} patterns and OTB \omcr{2}{values}.
\end{takeawaybox}

\subsection{\proposal on COTS HBM\jwcr{1}{2} chips}
\label{sec:hbm2}

\jwcr{1}{We extend \omcr{2}{our} analysis \omcr{2}{of \proposal} to real HBM2 chips.}
We test 3 HBM2 chips and analyze \threshold distribution
\jwcr{1}{across} 9216 rows \omcr{2}{(i.e., 3072 rows per HBM2 chip, sampled from three channels and two banks per channel).}
Fig.~\ref{fig:hbm} (left) shows the normalized \threshold values (y-axis) of the three HBM2 chips (hue) across different OTBs. Fig.~\ref{fig:hbm} (right) \jwcr{1}{compares} the \threshold distributions (y-axis) under \baseline and \proposal (hue) over the same OTB range. We make the following observation:

\begin{figure}[!ht]
    \centering
    \includegraphics[width=\linewidth]{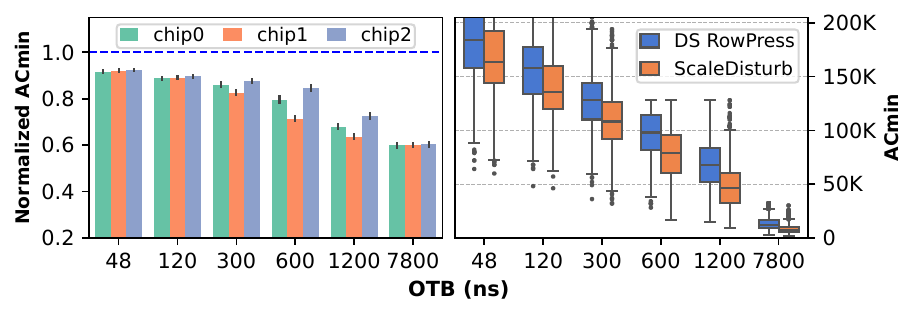}
    \caption{Normalized \threshold and \threshold distribution of HBM2 chips}
    \label{fig:hbm}
\end{figure}

\begin{obsvbox}
\proposal amplif\omcr{2}{ies} HBM2 chips' vulnerability to read disturbance.
\end{obsvbox}
We find that across all three HBM2 chips tested, \proposal decreases \threshold by an average (maximum) of 8\% (38\%), 11\% (51\%), 16\% (49\%), 22\% (58\%), 32\% (60\%), 40\% (63\%) than \baseline across the six evaluated OTBs. We expect our other observations for DDR4 chips ($\S$\ref{sec:ddr4_char} and $\S$\ref{sec:distinguish}) will hold for HBM2 chips as well because both \omcr{2}{DDR4 and HBM2 DRAM} use the same underlying DRAM cell array structure.
\begin{takeawaybox}
\proposal{} \jwcr{1}{widely affects} DRAM chips: not only DDR4 chips, but also HBM2 chips are vulnerable to \proposal.
\end{takeawaybox}

\subsection{Hypothetical Explanation for \proposal}
\label{sec:physical_explanation}

We provide a hypothesis \jwcr{1}{that could explain the \threshold reduction induced by \proposal}. Electron migration and injection into the victim cell are primary mechanisms driving DRAM read disturbance bitflips~\cite{ryu2017overcoming,yang2019trap,walker2021ondramrowhammer,zhou2023double,zhou2024understanding,zhou2024unveiling}. Prior work~\omcr{2}{\cite{zhou2024understanding,zhou2024unveiling}} shows that enhanced electric fields accelerate electron migration, which is the primary mechanism underlying RowPress-induced leakage. We \jwcr{1}{hypothesize that} the \jwcr{1}{\threshold} reduction caused by \proposal{} \jwcr{1}{could be attributed} to three factors: (i) Enhanced electric fields drive electron migration more effectively than trap-assisted leakage or crosstalk-induced mechanisms~\omcr{2}{\cite{zhou2024unveiling,zhou2024understanding}}. (ii) In double-sided patterns, the electric fields from the two aggressor wordlines can interact and partially cancel each other. (iii) \proposal{} introduces temporal asymmetry that breaks this interference, thereby increasing net electron migration. {We call for future device-level studies to develop a better
understanding of the inner workings of \proposal{}, just as device-level
studies~\cite{zhou2024unveiling,zhou2024understanding} did for RowPress after the RowPress paper \cite{luo2023rowpress,luo2024rowpress} demonstrated the RowPress phenomenon experimentally on real DRAM chips.}
\setcounter{version}{3}
\section{\proposal Sensitivity Study}
\jwcr{1}{We examine the sensitivity of \proposal bitflips to \omcr{2}{data pattern} (\secref{sec:data_pattern}) and \omcr{2}{temperature} (\secref{sec:temperature})}

\subsection{Data Pattern}
\label{sec:data_pattern}
We \jwcr{1}{evaluate} the impact of data pattern \omcr{2}{stored in DRAM rows} on \proposal. 
Fig.~\ref{fig:data_pattern} shows \threshold distribution (y-axis) across varied OTBs (x-axis) for two data patterns, \textit{Rowstripe0} and \textit{Rowstripe1} \hluocr{3}{(defined in~\secref{sec:method})}. We make the following observation:

\begin{figure}[!h]
    \centering
    \includegraphics[width=1.0\linewidth]{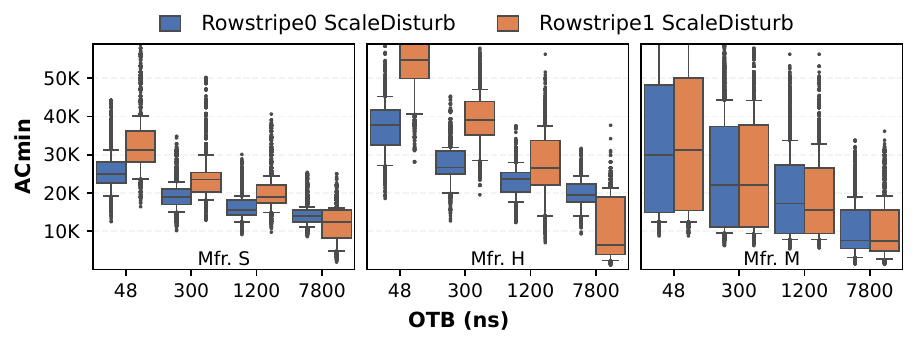}
    \caption{Distribution of \threshold under \proposal, with data patterns \textit{Rowstripe0} and \textit{Rowstripe1}}
    \label{fig:data_pattern}
\end{figure}

\begin{obsvbox}
{\textit{Rowstripe0} yields lower \threshold at small OTBs\omcr{3}{.} \textit{Rowstripe1} yields lower \threshold at large OTBs.}
\end{obsvbox}

\jwcr{1}{When OTB increases from 48\,ns to \qty{7.8}{\micro\second}, the \omcr{2}{ratios of} \threshold under \textit{Rowstripe1} relative to \textit{Rowstripe0} \omcr{2}{are} 1.27$\times$, 1.26$\times$, 1.21$\times$, and 0.83$\times$ for Samsung, 1.46$\times$, 1.43$\times$, 1.17$\times$ and 0.53$\times$ for SK Hynix, and 1.02$\times$, 1.02$\times$, 0.99$\times$, and 0.99$\times$ for Micron.}
This \omcr{2}{trend} \jwcr{1}{indicates a reversal in data pattern sensitivity: \textit{Rowstripe0} \omcr{2}{yields lower \threshold} at small OTBs, while \textit{Rowstripe1} \omcr{3}{yields lower \threshold} at larger OTBs.}
We hypothesize that \omcr{2}{when OTB is small, trap-assisted leakage~\cite{ryu2017overcoming,yang2019trap,walker2021ondramrowhammer,zhou2023double} is more effective than disturbance from enhanced electric field (due to prolonged row open times)~\cite{zhou2024understanding,zhou2024unveiling}}, consistent with the observation that $0\rightarrow1$ bitflips occur first. As OTB \omcr{2}{increases}, \omcr{2}{electric-field-induced disturbance becomes more effective at} inducing $1\rightarrow0$ bitflips.

\begin{takeawaybox}
{\proposal's sensitivity to d}ata pattern \omcr{2}{varies} with OTB: \textit{Rowstripe0} \omcr{2}{is more effective at inducing} 0$\rightarrow$1 \omcr{2}{bit}flips \omcr{2}{when OTB is small}, \textit{Rowstripe1} {becomes more effective at inducing} 1$\rightarrow$0 \omcr{2}{bit}flips as OTB \omcr{2}{increases}.
\end{takeawaybox}

\subsection{Temperature}
\label{sec:temperature}
\begin{figure}[!h]
    \centering
    \includegraphics[width=1.0\linewidth]{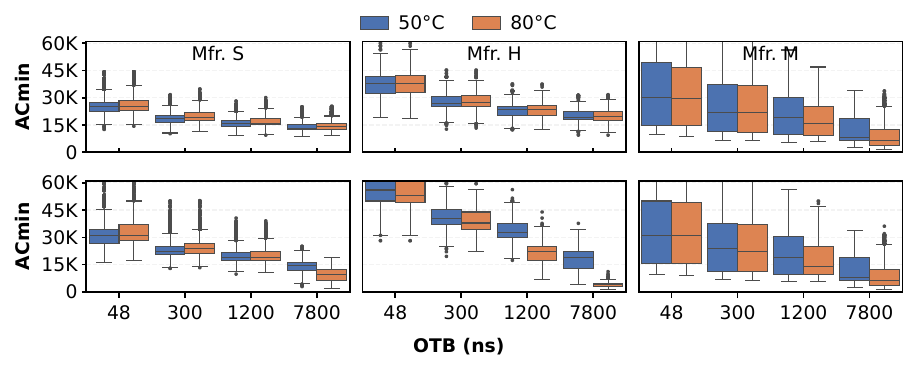}
    \caption{Effect of temperature (\qty{50}{\celsius} and \qty{80}{\celsius}) on \threshold of \proposal{} under \textit{Rowstripe0} (top) and \textit{Rowstripe1} (bottom).}
    \label{fig:temperature}
\end{figure}

We evaluate the impact of temperature on \jwcr{1}{the vulnerability of DRAM rows} under \proposal. Fig.~\ref{fig:temperature} shows the \threshold distributions (y-axis) across \omcr{2}{all tested} OTB values (x-axis) for two data patterns, \textit{Rowstripe0} (top) and \textit{Rowstripe1} (bottom), at \qty{50}{\celsius} and \qty{80}{\celsius} (hue), across \omcr{2}{all tested DRAM modules from} three manufacturers. \omcr{2}{Each box plot shows the distribution of \threshold across all tested rows: whiskers denote the minimum and maximum values, and the dashed line indicates the median.} \jwcr{1}{We make two key observations:}

\begin{obsvbox}
Under \textit{Rowstripe0}, \proposal shows limited temperature sensitivity at \omcr{2}{all} OTB \omcr{2}{values}.
\end{obsvbox}

\hluocr{3}{Under \textit{Rowstripe0}, a higher temperature of \qty{80}{\celsius} only slightly changes \threshold compared to \qty{50}{\celsius}. On average, at \qty{80}{\celsius}, \threshold changes by \param{+5.1\%}, \param{+0.6\%}, and \param{-11.2\%} for Samsung, SK Hynix, and Micron modules, respectively).}

\begin{obsvbox}
Under \textit{Rowstripe1}, higher temperature \hluocr{3}{significantly} \rbm{lowers} the \threshold at large OTB values.
\end{obsvbox}
\hluocr{3}{We observe that} \omcr{2}{under \textit{Rowstripe1}}, \hluocr{3}{a higher temperature of \qty{80}{\celsius}} \omcr{2}{decreases} the average {\threshold} by 31.0\%, 78.8\%, and 25.4\% for \hluocr{3}{Samsung, SK Hynix, and Micron modules}, respectively, \hluocr{3}{compared to \qty{50}{\celsius}}.

We hypothesize that higher temperature amplifies \omcr{2}{electric-field-induced disturbance (e.g., by accelerating electron migration in the electric field)~\cite{luo2023rowpress,zhou2024understanding,zhou2024unveiling}}, thereby \hluocr{3}{enhancing $1\rightarrow0$ leakage}. \hluocr{3}{We call for more device-level research to explain why \proposal's \threshold changes much less at a higher temperature of \qty{80}{\celsius} for the \textit{Rowstripe0} data pattern than for \textit{Rowstripe1}.}

\begin{takeawaybox}
\omcr{2}{Under \proposal access pattern,} \hluocr{3}{\threshold is less sensitive to a higher temperature for \textit{Rowstripe0} than for \textit{Rowstripe1} as OTB increases.} 
\end{takeawaybox}
\setcounter{version}{2}
\section{Real System Demonstration of \proposal}
We experimentally demonstrate that a simple user-level program can induce \proposal bitflips on a real DDR4-based system, despite the presence of periodic refresh and in-DRAM TRR\omcr{2}{~\cite{hassan2021utrr,frigo2020trrespass}} mechanisms. \omcr{2}{W}e describe our real system setup \jwcr{1}{(}\secref{sec:realsystem_setup}\jwcr{1}{)}, methodolog\omcr{2}{y} \jwcr{1}{(}\secref{sec:realsystem_method}\jwcr{1}{)}, and testing results \jwcr{1}{(}\secref{sec:realsystem_results}\jwcr{1}{)}.

\subsection{Experimental Setup}
\label{sec:realsystem_setup}
\noindent\textbf{System Configuration.} We \jwcr{1}{conduct our experiments on a system running} Ubuntu 18.04 with Linux kernel 5.4.0-131-generic~\cite{launchpad-linux-5.4.0-131.147}. \jwcr{1}{The system uses} an Intel i5-10400 Comet Lake processor~\cite{intel-core-i5-10400} and a 16\,GB dual-rank DDR4 DRAM module~\cite{samsung-8gb-c-die-dimm} \jwcr{1}{with} target row refresh (TRR)~\cite{frigo2020trrespass,hassan2021utrr} as an in-DRAM mitigation mechanism.

\noindent\rbc{\textbf{Memory Address Mapping.}~We reverse engineer the processor's address mapping from physical memory address to DRAM rank, bank, row, and column addresses using the DRAMA \jwcr{1}{method}~\cite{pessl2016drama} and \jwcr{1}{its extensions in} prior work\omcr{2}{~\cite{de2021smash,frigo2020trrespass,luo2023rowpress,luo2024rowpress,meyer2025phoenix}}. First,
\jwcr{1}{accessing different rows within a bank (\emph{row conflicts}) incurs higher access latency than accessing the same row (\emph{row hits})}. We exploit \jwcr{1}{this latency difference to} identify address sets that map to the same bank but different rows. Second, prior work~\cite{seaborn2015physical,pessl2016drama} reports that Intel’s DRAM addressing functions exhibit linearity (e.g., an XOR
of many physical address bits). \jwcr{1}{We use
the identified address sets to reconstruct the addressing
functions.} \jwcr{1}{Specifically, w}e allocate a 1GB page using hugepage support~\cite{linux-hugetlbpage-support} to directly manipulate the least significant 30 physical address bits that encompass all DRAM rank, bank, and row address bits. We generate pointers to aggressor and victim rows within \jwcr{1}{the} page to place them in physically adjacent DRAM rows.}

\subsection{\proposal on \omcr{2}{a} Real System}
\label{sec:realsystem_method}
\noindent\textbf{Challenges and Methodolog\omcr{2}{y}.} Demonstrating \proposal bitflips on real systems \jwcr{1}{presents} two key challenges. \jwcr{1}{First,} asymmetrically \jwcr{1}{varying} the open times of the two aggressor rows is difficult on real systems \jwcr{1}{because the command sequences are scheduled by} the memory controller, \omcr{2}{over which software} does not have direct control. Prior work~\cite{luo2023rowpress} \jwcr{1}{keeps the DRAM row open for a long duration} by repeatedly accessing different cache blocks in the same DRAM row. \jwcr{1}{This approach exploits the memory controller scheduling policy, which prioritizes memory requests targeting the currently open row during scheduling}\omcr{2}{~\cite{rixner2000frfcfs,mutlu2008parbs,mutlu2016bliss}}. Building on this, we \jwcr{1}{create} asymmetric row open times by accessing different numbers of cache blocks in the two aggressor rows. 
\jwcr{1}{Second,} TRR mechanisms can detect aggressor rows {in a \proposal access pattern and prevent us from inducing bitflips by refreshing the victim rows}. \jwcr{1}{However,} TRR \jwcr{1}{mechanisms are typically limited by} tracking capacity, \jwcr{1}{and can be bypassed by certain access patterns that access many other dummy aggressor rows (dummy rows) besides the real aggressor rows~\cite{frigo2020trrespass,hassan2021utrr}}. Such access patterns use dummy rows to \jwcr{1}{exhaust} TRR's tracking capacity, so that the real aggressor rows remain undetected.

\noindent\textbf{Test Program.} 
Algorithm~\ref{alg:real-system} shows the \jwcr{1}{key part} of our test program. We \jwcr{1}{\circledcharblack{1}}initialize the rows using a checkerboard data pattern \omcr{2}{that} induces the highest average bit error rate (BER) on DDR4 chips~\cite{kim2020revisiting}~\rbi{(line\omcr{2}{s} 2-3),} \circledcharblack{2}issue \jwcr{1}{memory requests} targeting different cache blocks within the same aggressor row to keep the row open for a long time~(i.e., \texttt{NUM\_READ1} for $aggressor1$ (line 8), and \texttt{NUM\_READ2} for $aggressor2$ (line 12)), \circledcharblack{3}flush the cache blocks to DRAM to ensure that subsequent \jwcr{1}{memory requests bypass the processor caches and go to DRAM}~(line\omcr{2}{s} 10,\,14), \circledcharblack{4}alternately access two aggressor rows in an activation iteration to change the open row \rbi{(line\omcr{2}{s 9,\,13})} as they are in the same bank, \circledcharblack{5}execute a\omcr{2}{n} \emph{mfence} instruction \rbi{(line 16)} to ensure that data \omcr{2}{is} fully flushed~\cite{kim2014flipping}, and \circledcharblack{6}access 16 dummy rows,~\jwcr{1}{four times each,} to bypass TRR~\rbi{(line 18)}. 
For every victim row, we \circledcharblack{7}execute
this access pattern for~\rbm{800\,K} iterations~\rbi{(line 6)} to gather statistically significant results \jwcr{1}{and record the bitflips in the victim row}~\rbi{(line 20)}. \rbi{\revci{D1.4}We use \texttt{NUM\_AGGR\_ACTS} to regulate the activation frequency of dummy rows. \jwcr{1}{A smaller} \texttt{NUM\_AGGR\_ACTS} increases the dummy row activation frequency, increasing the likelihood that aggressor rows evade TRR detection.}

\begin{algorithm}[h]
\centering
\captionsetup{font=footnotesize}
\caption{\textbf{Pseudocode of Real System Test Program}}
\label{alg:real-system}
\begin{minipage}{0.92\linewidth}
\begin{algorithmic}[1]
\scriptsize
\State $aggressor1, aggressor2 = \mathrm{find\_aggressor\_rows}(victim)$
\State $\mathrm{initialize}(victim, 0x55555555)$
\State $\mathrm{initialize}(aggressor1, aggressor2, 0xAAAAAAAA)$
\For{$\mathrm{NUM\_READ1} = 0$ to $\mathrm{TOTAL\_READ}$}
    \State $\mathrm{NUM\_READ2 = TOTAL\_READ - NUM\_READ1}$
    \For{$iter = 1$ to $\mathrm{NUM\_ITER}$}
        \For{$i = 1$ to $\mathrm{NUM\_AGGR\_ACTS}$}
            \For{$j = 1$ to $\mathbf{NUM\_READ1}$}
                \State $\ast aggressor1[j]$
                \State $\mathrm{clflushopt}(aggressor1[j])$
            \EndFor
            \For{$j = 1$ to $\mathbf{NUM\_READ2}$}
                \State $\ast aggressor2[j]$
                \State $\mathrm{clflushopt}(aggressor2[j])$
            \EndFor
            \State $\mathrm{mfence}()$
        \EndFor
        \State $\mathrm{activate\_dummy\_rows}()$
    \EndFor
    \State $\mathrm{record\_bitflips}[victim] = \mathrm{check\_bitflips}(victim)$
\EndFor
\end{algorithmic}
\end{minipage}
\end{algorithm}

\noindent\textbf{\rbm{Experimental} Parameters.} We {evaluate} \texttt{NUM\_AGGR\_ACTS} $\in$ \{2, 3, 4\} and \texttt{TOTAL\_READS} $\in$ \{16, 32, 64, 128\} on 1500 arbitrarily selected victim rows.

\subsection{Real System Testing Result\jwarv{2}{s}}
\label{sec:realsystem_results}

\begin{figure}[h]
    \centering    
    \includegraphics[width=1.0\linewidth]{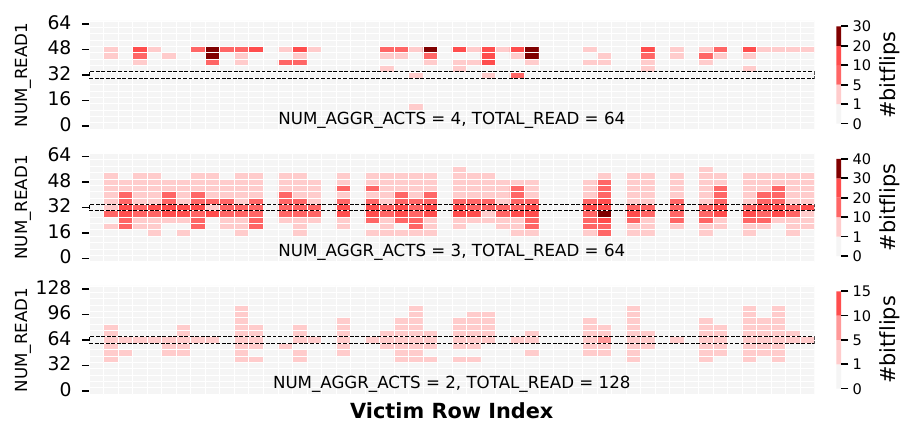}
    \caption{Bitflip counts across rows under different \texttt{NUM\_AGGR\_ACTS} while sweeping \texttt{NUM\_READ1} \omcr{2}{for} different \omcr{2}{\texttt{TOTAL\_READ}} values.}
    \label{fig:heatmap}
\end{figure}

Fig.~\ref{fig:heatmap} shows \jwcr{1}{the number of bitflips} across 50 rows \omcr{2}{with the highest bitflip counts among the 1500 tested rows, using} three stacked subplots for \jwcr{1}{\texttt{NUM\_AGGR\_ACTS} = \{4, 3, 2\}} (top to bottom). The x-axis \jwcr{1}{denotes} the row index, and the y-axis \jwcr{1}{denotes} \texttt{NUM\_READ1}, from 0 to \texttt{TOTAL\_READ}. Bitflips are shown with a color gradient (darker = more, grey = none). Dashed boxes \jwcr{1}{denote} double-sided RowPress cases \jwcr{1}{where \texttt{NUM\_READ1} = \texttt{NUM\_READ2}. We make the following observation:}

\begin{obsvbox}
\proposal{} \omcr{2}{induces bitflips in cells where} double-sided RowPress \omcr{2}{cannot}, \omcr{2}{under} low dummy \omcr{2}{row} activation frequency.
\end{obsvbox}

\proposal flips \omcr{2}{cells in} 33 of 50 rows versus 3 for double-sided RowPress under \texttt{NUM\_AGGR\_ACTS} = 4. \jwcr{1}{For the three rows flipped by both access patterns}, \proposal induces up to 34 bitflips, compared to at most 5 for \jwcr{1}{double-sided} RowPress.

\begin{figure}[h]
    \centering
    \includegraphics[width=\linewidth]{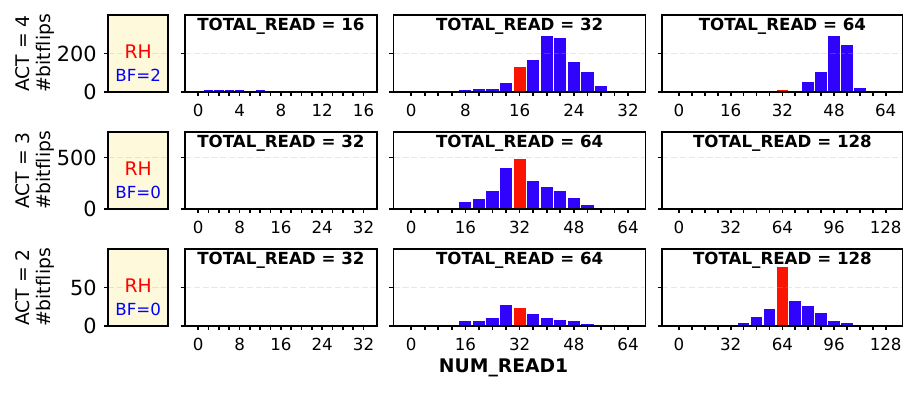}
    \caption{Number of bitflips under \proposal (blue), RowHammer (light-yellow), and double-sided RowPress (red)}
    \label{fig:real_num_bitflips}
\end{figure}

\jwcr{1}{Fig.}~\ref{fig:real_num_bitflips}\jwcrcomment{I would keep the ACT at left side for figure 17, because a row of subplots share the same ACT value and it would be too cluttered if I include it inside each subplot} shows the \jwcr{1}{bitflips distribution} across \jwcr{1}{\texttt{NUM\_AGGR\_ACTS}} (shown as ACT \omcr{2}{on the y-axis labels}) and \texttt{TOTAL\_READ}. Each subplot {shows the number of} bitflips (y-axis) versus \texttt{NUM\_READ1} (x-axis), with blue and red bars for \proposal and double-sided RowPress, \jwcr{1}{respectively}, and RowHammer with yellow backgrounds. \jwcr{1}{We make the following observation:}

\begin{obsvbox}
\proposal{} \omcr{2}{induces bitflips in} more cells than double-sided RowPress\omcr{2}{, under} low dummy row activation frequency.
\end{obsvbox}

\jwcr{1}{We observe that the total bitflip \omcr{2}{count} varies with \texttt{NUM\_READ1} under different dummy row activation frequencies. At a low dummy activation frequency (ACT\,=\,4), \proposal consistently induces more bitflips than \baseline: 289 vs.\ 8, 290 vs.\ 130, and 10 vs.\ 6 for \texttt{TOTAL\_READ} = 64, 32, and 16, respectively.}
Only 2 RowHammer bitflips are observed at \texttt{NUM\_AGGR\_ACTS} = 4, and none at lower \texttt{NUM\_AGGR\_ACTS} counts.

We conclude that on a real DDR4-based Intel system with \jwcr{1}{in-DRAM} TRR \jwcr{1}{mechanism\omcr{2}{s}}, our user-level test program demonstrates that \proposal: (1) induce\omcr{2}{s} bitflips in scenarios where double-sided RowPress \jwcr{1}{cannot}, (2) induces significantly more bitflips than double-sided RowPress \rbm{at} low dummy row activation frequency, and (3) consistently induces more bitflips than RowHammer. These findings suggest that read disturbance attacks on real systems (e.g.,~\cite{frigo2020trrespass,hassan2021utrr}) can be made more effective by incorporating \proposal.

\begin{takeawaybox}
A user-level program leveraging \proposal 1) induce\omcr{2}{s} bitflips in cases where \baseline cannot, or induce more when both \omcr{2}{can}, 2) \jwcr{1}{consistently} induces more bitflips than RowHammer.
\end{takeawaybox}

\section{Implications and Mitigations}
The key findings \atb{of} our \jwcr{1}{real chip} characterization \atb{and real system evaluations} have critical implications for the security and reliability of
(1) existing academic and industrial read disturbance mitigation techniques~\omcr{2}{\mitigatingRowHammerAllCitations{}}
and (2) standardized \jwcr{1}{mitigation} mechanisms already deployed in real systems\omcr{2}{~\cite{kim2014flipping,kim20231ddr5,bennett2021panopticon,tomita2022extracting,jedec2024ddr5,canpolat2024understanding,canpolat2025chronus}}. These mitigation \atb{techniques} fundamentally rely on accurately profiling the \atb{read disturbance} threshold (\threshold) of each DRAM row using existing \atb{memory} access patterns \atb{(e.g., double-sided RowPress\omcr{2}{~\cite{luo2023rowpress,luo2024rowpress}} \jwcr{1}{and RowHammer\omcr{2}{~\cite{kim2014flipping, kim2020revisiting,luo2024experimental}}})}. A lower \threshold, if undetected, can lead to insufficient protection\omcr{2}{, and thus bitflips}. Our results show that \proposal can reduce the \threshold of a DRAM row \omcr{2}{to a value that is} up to \maxavr{} \jwcr{1}{than} existing \jwcr{1}{memory} access patterns. \atb{Thus, a read disturbance mitigation technique configured with \threshold values
{obtained} using existing memory access patterns~\nb{might} \emph{not} securely mitigate read disturbance bitflips.}
 
\rbm{We} examine \omcr{2}{four} potential \jwcr{1}{solutions} to mitigate \proposal bitflips\rbm{:} Error Correcting Codes\jwcr{1}{~(\secref{sec:ecc})}, applying safety margins~\jwcr{1}{(\secref{sec:safety_margin})}, \omcr{2}{adaptive} existing RowPress mitigations~\jwcr{1}{(\secref{sec:rowpress_m})}, and a new mitigation mechanism, \sol, to effectively mitigate \proposal{} \jwcr{1}{bitflips} with low overhead~\jwcr{1}{(\secref{sec:teacup})}.

\subsection{Error Correcting Codes (ECC)}
\label{sec:ecc}

\begin{figure}[htb]
\vspace{0.5em}
    \centering
    \includegraphics[width=\linewidth]{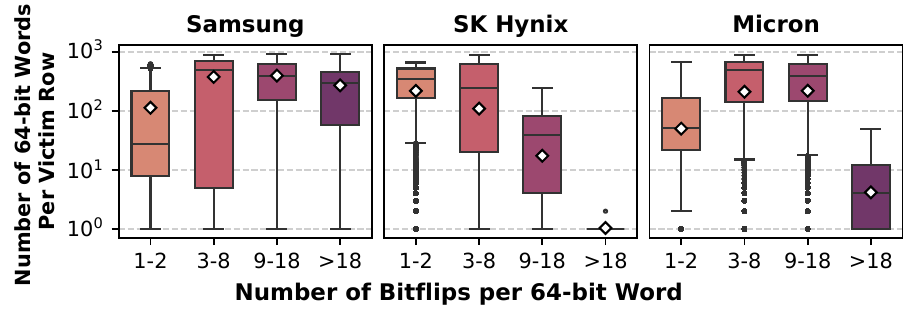}
    \caption{Distribution of 64-bit words by bitflip count across three modules under \proposal (\textit{Rowstripe1}, 80 °C, 60 ms).}
    \label{fig:ecc_stat}
\end{figure}

We examine \jwcr{1}{ECC's} error-correcting capability in mitigating \proposal{} \jwcr{1}{bitflips by analyzing} the number of bitflips induced by \proposal in each 64-bit word across a wide range of OTBs. \jwcr{1}{For each victim row}, we \jwcr{1}{activate} the aggressor rows as many times as possible within 60 ms at 80°C using \textit{Rowstripe1} data pattern. 
Fig.~\ref{fig:ecc_stat} shows the distribution of \jwcr{1}{the number of 64-bit words per victim row (y-axis) across different error-intensity categories (x-axis) for} three DRAM modules, one from each manufacturer.

We make two key observations. First, \proposal{} \jwcr{1}{can induce up to 40 bitflips (not shown) in a 64-bit word, far exceeding the correction capability of widely used ECC techniques such as SECDED~\cite{hamming1950error} and ChipKill~\cite{dell1997white,locklear2000chipkill,memoryadvanced}}\omcr{2}{.}\footnote{Chipkill~\cite{dell1997white,locklear2000chipkill,memoryadvanced} can correct any errors \omcr{2}{(i.e., one-symbol errors)} from a single DRAM device and detect two-symbol errors. \jwcr{1}{Because we observe up to 40 bitflips in a 64-bit word, at least five symbols (i.e., data from five DRAM devices for x8 chips) will be erroneous. Thus,} Chipkill cannot provide guaranteed mitigation against \proposal.}
\jwcr{1}{A (7,4) Hamming code~\cite{hamming1950error} can correct such bitflips with very large DRAM storage overheads (75\%,
i.e., three parity bits for every four data bits)}. Thus, relying on ECC \jwcr{1}{alone} to \jwcr{1}{prevent} \proposal{} \jwcr{1}{bitflips} is \jwcr{1}{a \omcr{2}{very} expensive solution}.
Second, a large fraction of words exhibit at least three bitflips across the three modules\jwcr{1}{: 55.0}\%, \jwcr{1}{25.8}\%, and \jwcr{1}{52.7}\%, respectively. This {limits the effectiveness of} \omcr{2}{page offlining (or} \textit{Memory Page Retirement}\omcr{2}{)} techniques~\cite{meza2015revisiting,tang2006assessment}, which forcibly offline faulty pages, because retiring pages mapped to rows with multiple faulty bits \jwcr{1}{can substantially} reduce usable capacity, thereby increasing storage overhead by \jwcr{1}{55.0}\% \omcr{2}{based on our results}.

\subsection{Using a Safety Margin for \proposal}
\label{sec:safety_margin}
\label{sec:overhead}
Read disturbance mitigation mechanisms~\mitigatingRowHammerAllCitations{} fundamentally rely on a \omcr{2}{protection} \emph{threshold}, measured as the minimum \threshold observed across all tested rows within a \jwcr{1}{DRAM} module,
to trigger preventive \jwcr{1}{operations (e.g., preventive refreshes) that} avoid bitflips. To \jwcr{1}{prevent bitflips that} \proposal{} \rbm{induces at a lower \threshold}, system designers \rbi{can lower the configured threshold by a safety \revci{B3.2}margin (e.g., 10\%)}.\jwcrcomment{The 10\% here is an example to let readers better understand what the safety margin is; we have a trade-off analysis between security and overhead in the part of performance overhead.}
To \jwcr{1}{quantify the performance and energy overhead} of {applying} safety margins, we evaluate five state-of-the-art RowHammer mitigation mechanisms: 1) Graphene~\cite{park2020graphene}, a deterministic mechanism that uses the Misra-Gries frequent-item counting algorithm~\cite{misra1982finding} and maintains frequently accessed row counters
\jwcr{1}{entirely} within the memory controller, 2) Hydra~\cite{qureshi2022hydra}, a deterministic mechanism
that \jwcr{1}{stores row} counters in the DRAM chip and caches them
in the memory controller, 3) PRAC~\cite{canpolat2024understanding}, \jwcr{1}{a deterministic mechanism that stores a counter for each row in the DRAM chip}, 4) Periodic RFM (PRFM)~\cite{jedec2024ddr5}, where the memory controller issues an RFM command \jwcr{1}{once the activation count to a bank} reaches a predefined threshold (\textit{RFMth}), with no back-off signal from the DRAM chip, and 5) PARA~\cite{kim2014flipping}, a memory-controller-based mechanism that \jwcr{1}{probabilistically refreshes potential victim rows on each activation, without counters}. We perform our evaluation on a DDR5-based system modeled \rbm{in} Ramulator 2.0~\cite{luo2023ramulator2,ramulator2github} \omcr{2}{(successor of Ramulator~\cite{kim2016ramulator,ramulatorgithub})}.\footnote{We use 60 {four-core} benign workloads, {each combining four single-core benchmarks from a pool of 57} drawn from SPEC CPU2006~\cite{spec2006}, SPEC CPU2017~\cite{spec2017},~TPC~\cite{tpcbenchmarks}, MediaBench~\cite{fritts2009media}, and YCSB~\cite{ycsb}. 
Each workload is {memory-intensive, with} a last-level cache misses-per-kilo-instructions (MPKI) $\geq$ 20. We configure the simulator with four 4\,GHz out-of-order cores, dual-rank DDR5 DRAM, FR-FCFS scheduling and
open-row policy.}

\begin{figure}[h]
    \centering
    \includegraphics[width=1.0\linewidth]{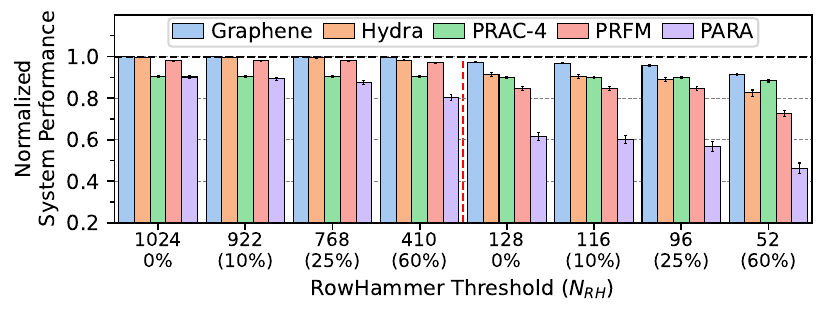}
    \caption{Performance (normalized IPC) of five read disturbance mitigation mechanisms under 60 four-core, memory-intensive workload mixes, evaluated at two baseline thresholds (1024 and 128), each with safety margins of 0\%, 10\%, 25\%, and 60\%.}
    \label{fig:overhead}
\end{figure}

\noindent\textbf{Performance Overhead.}~\figref{fig:overhead} shows the performance (y-axis) of \jwcr{1}{the} five\jwcrcomment{no ABACus and CoMet? Added them to GDoc} \nb{RowHammer} mitigation mechanisms (hue), normalized to a baseline system \textit{without} \nb{any RowHammer} mitigation \nb{mechanism}. The x-axis presents two baseline \jwcr{1}{thresholds} (1024 and 128), each evaluated with three safety margins: 10\%, 25\%, and 60\% (e.g., a \jwcr{1}{baseline} threshold of 128 with a 60\% safety margin reduces to $\lceil 128 \times (1-0.6) \rceil = 52$).\footnote{These \rbm{safety} margins are derived from empirical characterization: an average \threshold reduction around 10\% (9.6\%) across all tested rows;~\atb{a 25\% average \threshold reduction across rows in one tested DIMM (that yields the highest \threshold reduction on average across its rows among all tested DIMMs);}~a worst-case reduction of up to 63\% observed in \atb{one} victim row.}

\atb{We make two key observations.}~\textbf{First}, large safety margins lead to substantial performance degradation. For example, relative to the \emph{same} mitigation mechanisms with 0\% safety margin, \rbc{applying a 60\% margin at threshold 128 \omcr{2}{(i.e., configuring the mechanism for a threshold of 52)} reduces system performance by \param{6.1\%}, \param{10.0\%}, \param{2.0\%}, \param{14.9\%}, and \param{28.6\%} for Graphene, Hydra, PRAC, PRFM, and PARA, respectively.}
\textbf{Second}, \omcr{2}{a} small safety margin incur\omcr{2}{s small additional} overhead. 
\omcr{2}{Applying a 25\% margin at threshold 128 reduces system performance by 1.6\%, 2.8\%, 0.0\%, 0.0\%, 9.3\% for Graphene, Hydra, PRAC, PRFM, and PARA, respectively, compared to the same mitigation mechanisms without a safety margin.}
However, this comes at the risk of inducing bitflips in devices that experience bitflips at lower~\threshold as shown by our empirical analysis ($\S$\ref{sec:ddr4_char}). We conclude that applying a safety margin introduces an inherent security–performance trade-off: larger margins provide stronger protection but also incur substantial performance overhead.

\begin{figure}[h]
    \centering
    \includegraphics[width=1.0\linewidth]{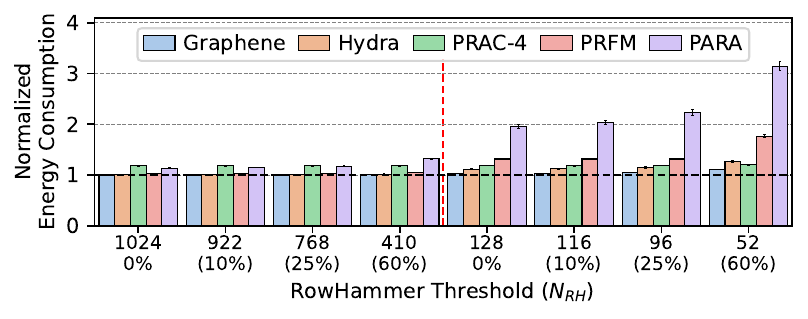}
    \caption{Energy impact of evaluated read disturbance mitigation
            mechanisms, evaluated at two baseline thresholds with three margins.}
    \label{fig:energy}
\end{figure}

\noindent\textbf{Energy Overhead.} We evaluate the impact of \proposal on DRAM energy consumption using Ramulator 2.0\omcr{2}{~\cite{ramulator2github,luo2023ramulator2}}. We track key DRAM commands 
and account for background energy. The energy is then calculated using \omcr{2}{{DRAMPower}~\cite{drampower5}}. \figref{fig:energy} shows the energy consumption of \jwcr{1}{the} five mitigation mechanisms normalized to a baseline \jwcr{1}{system} without \jwcr{1}{any} mitigation \jwcr{1}{mechanisms}. 

We observe that using large safety margins increases the energy consumption {by} \param{12.3\%} on average (up to \param{58.8\%}) \omcr{2}{across all evaluated mitigation mechanisms}. \rbm{A}t \nb{a} threshold \nb{of} 128 with a 60\% safety margin, energy consumption increases by \param{7.4\%}, \param{13.2\%}, \param{1.5\%}, \param{33.2\%}, \param{58.8\%} for Graphene, Hydra, PRAC, PRFM, and PARA, respectively.

We conclude that securely mitigating \proposal bitflips by {applying} large safety margins to RowHammer mitigation mechanisms \jwcr{1}{incurs} significant performance and energy overheads.

\subsection{\omcr{2}{Adapting} Existing RowPress Mitigation\omcr{2}{s}}
\label{sec:rowpress_m}
We further \atb{discuss} the effectiveness of two RowPress mitigation mechanisms against \proposal.
Luo et al.~\cite{luo2023rowpress} propose \jwcr{1}{an approach} that \jwcr{1}{mitigates} RowPress bitflips by 1) \jwcr{1}{enforcing a memory-controller-side limit on the maximum row open time (\emph{tMRO})}, and 2) setting the protection threshold to the corresponding \threshold within that \emph{tMRO}. 
This approach is {limited} against \proposal \omcr{2}{because} enforcing a \emph{t{MRO}} limit reduces the row hit rate, degrading the performance of applications with strong \jwcr{1}{memory} locality\omcr{2}{\cite{rixner2000frfcfs,mutlu2008parbs,mutlu2016bliss}}.

Saxena et al.~\cite{saxena2024impress} propose \emph{ImPress}, which mitigates RowPress bitflips by mapping each activation's open time to counter increments: an activation that holds a row open for $\textit{N} \times$ \tRC{} {is counted} as \textit{N} consecutive RowHammer \jwcr{1}{activations}, \jwcr{1}{incrementing the row's activation counter} by \textit{N}.
Because ImPress does not enforce a \emph{tMRO} limit, it avoids the additional activations from forced row closures and is compatible with in-DRAM counter mechanisms.
However,~\emph{ImPress} is challenged by the temporal asymmetry of \proposal: 
one of the two aggressor rows accumulates counter increments faster than the other, which can cause its counter to cross the protection threshold prematurely~\omcr{2}{(i.e., before its actual disturbance level becomes sufficient to induce bitflips)}, triggering unnecessary preventive refreshes.
For example, we define the \emph{tAggON ratio} as the ratio of the open times of the two aggressor rows under \proposal. \jwcr{1}{Consider a fixed total open time of $100 \times \textit{tRC}$ across both aggressor rows: they receive the same number of activations but with different row open times. Because \emph{ImPress} maps each activation's open time to counter increments, the resulting counter values differ}: a symmetric ratio (1:1) yields 50 counter increments per row, whereas an asymmetric ratio (3:1) yields 75 increments for \omcr{2}{one aggressor row with longer open time (faster row)} and 25 for \omcr{2}{the other aggressor row with shorter open time (slower row)}.
\jwcr{1}{The faster row's counter reaches the protection threshold before its physical disturbance level poses a real bitflip risk}, triggering \emph{unnecessary} preventive refreshes. These refreshes reduce bank availability and memory bandwidth, degrading system performance. \omcr{2}{To avoid such issues, we develop a new mitigation mechanism \sol, described in \secref{sec:teacup}}

\subsection{\rbc{Temporal Asymmetry-aware Counter Update}\revcc{CQ3}}
\label{sec:teacup}

\begin{figure}[h]
    \centering
    \includegraphics[width=\linewidth]{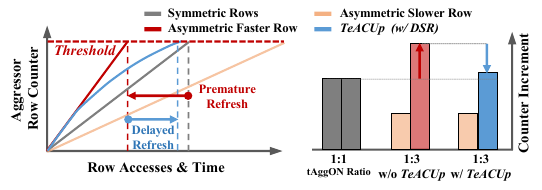}
    \caption{\sol overview: \omcr{2}{\sol reduces unnecessary preventive refreshes caused by temporal asymmetry by slowing the counter growth of the faster aggressor row using DSR.}}
    \label{fig:teacup_overview}
\end{figure}

\omcr{2}{W}e propose \emph{TeACUp} (Temporal-\omcr{2}{A}symmetry-\omcr{2}{a}ware Counter Update), \jwcr{1}{a mechanism that \omcr{2}{safely} slows the \omcr{2}{counter increment} of the faster aggressor row \omcr{2}{(defined in \secref{sec:rowpress_m})}, preventing it from crossing the protection threshold prematurely\jwcrcomment{defined in Section 7.3}}.~Specifically, \sol introduces a \emph{dynamic scale ratio} (\emph{DSR}), defined as $\mathrm{\textit{DSR}} = \min(\textit{AC}_1,\textit{AC}_2)/\max(\textit{AC}_1,\textit{AC}_2)$, where $\mathrm{\textit{AC}_1}$ and $\mathrm{\textit{AC}_2}$ are the {current counter value} of the two aggressor rows. For an activation with open time $N \times tRC$, \sol updates the aggressor counters as $\Delta C_{\text{faster}}=\max\{1,\lfloor N\!\cdot\!\mathrm{\textit{DSR}}\rfloor\}$ or $\Delta C_{\text{slower}}=N$, where \emph{DSR~}$\leq1$ downscales the faster row to balance the counter growth.
\omcr{2}{\jwcr{1}{Fig.~\ref{fig:teacup_overview} \omcr{2}{(left)} shows the counter increments of the two aggressor rows under \proposal. Due to asymmetric row open times, the faster aggressor row (red) accumulates counter increments more rapidly than both the slower aggressor row (yellow) and rows with symmetric row open times (gray). As a result, the faster row reaches the protection threshold earlier than the symmetric baseline, where the counter value accurately reflects the underlying disturbance level, leading to unnecessary preventive refreshes.
\sol reduces such refreshes by slowing the counter growth of the faster aggressor row (blue)}. \figref{fig:teacup_overview}\jwcrcomment{figure 21 replotted: changed "Saved" to "Delayed Refresh": delay the timing when the faster row crosses the protection threshold (thus trigger preventive refresh)} (right) shows an example how \sol slows the counter increment of the faster row using DSR.}

\noindent{\textbf{Security Analysis.} 
\sol guarantees secure protection as long as DRAM is accurately profiled and row activations are correctly mapped into counter increments, ensuring mitigation is invoked before any row exceeds the protection threshold.
\sol safely scales down the faster row's counter increment while preserving the counter value as an upper bound on activations, ensuring timely mitigation. In the symmetric case (\textit{DSR} = 1), the counter matches activations exactly. When \textit{DSR} $<$ 1, \sol scales down the counter increment for the current activation while preserving the accumulated counter value as an upper bound on the true activation count (\omcr{2}{gray line} in ~\figref{fig:teacup_overview}).

\noindent\textbf{Evaluation.}~We use PRAC as \jwcr{1}{the row-counter mechanism} for both the baseline \emph{ImPress}~\cite{saxena2024impress} and \sol. The memory controller \jwcr{1}{is configured} to issue four consecutive RFM commands upon receiving a back-off signal (i.e., ABO\omcr{2}{\cite{canpolat2025chronus,canpolat2024understanding}}). We \jwcr{1}{evaluate both mechanisms} using the workloads \rbm{described} in~\secref{sec:safety_margin}.
~\figref{fig:eval_impress_vs_teacup} shows the performance of \sol normalized to \emph{ImPress}. 
\jwcr{1}{Overall}, \sol improves system performance by 3.2\% on average (up to \param{9.2\%}).
\jwcr{1}{The performance improvement comes from }suppressing excessive growth in fast-row counters, thereby delaying them from reaching the \emph{ABO threshold} and reducing the number of RFM commands and preventive refreshes.
\sol{}'s effectiveness depends on the distribution of accesses across hot rows. When accesses are highly skewed and hot rows exhibit large differences in access duration, \sol exploits \emph{DSR} to slow fast-row counter growth, reducing RFMs and preventive refreshes. In contrast, when accesses are sparse or more evenly distributed, opportunities to reduce refreshes are limited, resulting in smaller performance gains.

\begin{figure}[h]
    \centering
    \includegraphics[width=1.0\linewidth]{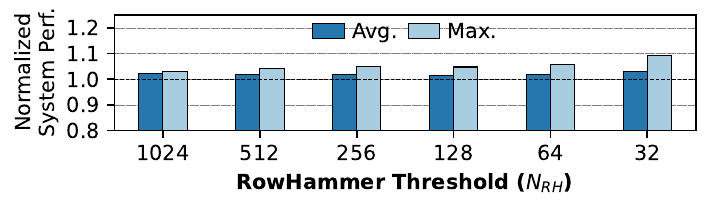}
    \caption{Performance evaluation: \sol normalized to ImPress}
\label{fig:eval_impress_vs_teacup}
\end{figure}

\noindent{\textbf{Overhead of Storage and Latency.}~\sol{} is compatible with counter-based mitigation mechanisms\jwcrcomment{No PRAC comparison? We use PRAC as a row-counter technique only. Will add it in the extended version} (e.g., PRAC\omcr{2}{~\cite{canpolat2025chronus,canpolat2024understanding}}) and requires no additional storage area for activation counters. Recent work~\cite{canpolat2025chronus} exploits subarray-level parallelism to \omcr{2}{hide} counter update \omcr{2}{latencies}. This design can be adopted in \sol to minimize the latency overhead of \emph{DSR} calculation, enabling concurrent multiple counter queries during bank precharge operations, thereby effectively hiding the associated latency.}

\setcounter{version}{3}
\section{Related Work}
To our knowledge, this is the first work to~\omcr{2}{introduce a new \omcr{3}{read disturbance} access pattern that employs temporal asymmetry, called \proposal{}, and analyzes real DRAM chip\omcr{3}{'}s read disturbance vulnerability with this pattern}. In this section, we discuss other related works.

\noindent
\textbf{Experimental DRAM Read Disturbance Characterization.} 
Existing DRAM read disturbance characterization works test either {only} RowHammer patterns\omcr{2}{~\cite{kim2014flipping, kim2020revisiting, orosa2021deeper, yaglikci2022understanding,lang2023blaster,he2023whistleblower,hassan2021utrr,park2016experiments}} or {separate RowHammer and RowPress patterns}\omcr{2}{~\cite{orosa2021deeper, luo2023rowpress, luo2024rowpress,luo2025revisiting, park2016experiments, park2016statistical, nam2024dramscope, nam2023xray,yaglikci2024svard,olgun2023hbm, olgun2024hbm, olgun2025vrd,yaglikci2022understanding,tugrul2025understanding,luo2026dejavu,yuksel2025pudhammer}}. A prior work~\cite{luo2024experimental} studies a combined RowHammer and RowPress pattern where one of the aggressor row in the double-sided pattern has the minimal \tagg{} (i.e., hammered) and the other has a longer \tagg{} (i.e., pressed). 
Many-sided RowHammer access patterns~\cite{kogler2022half,lang2023blaster,frigo2020trrespass,de2021smash} show that multiple aggressor rows can collaboratively induce \omcr{3}{bitflips in victim rows that are spatially distant from the aggressor rows.}
\jwcr{1}{They do \emph{not} study how read disturbance bitflips change when the \tagg{} value of both aggressor rows vary for a given \omcr{2}{Open} Time Budget as \proposal{} does.}

\noindent
\textbf{Device-Level Studies of DRAM Read Disturbance.}
Prior works on device-level mechanisms of RowHammer~\cite{yang2016suppression, park2016experiments, ryu2017overcoming, yang2019trap, walker2021ondramrowhammer, zhou2023double, li2023device-understanding} and RowPress~\cite{zhou2024understanding, zhou2024unveiling} do \emph{not} study the \proposal access pattern.~\omcr{2}{We believe new device-level studies are needed to properly understand \proposal{} effects (\secref{sec:physical_explanation}).}

\section{Conclusion}
\rbm{W}e \rbm{present} a new DRAM access pattern, \emph{\proposal}, \jwcr{1}{which amplifies DRAM's vulnerability to read disturbance by scaling the open time of two aggressor rows unevenly}. Our detailed characterization of \proposal on~\chips COTS DDR4 and 3 HBM2 DRAM chips shows that {\proposal} induce\omcr
{2}{s} bitflips in the victim row with fewer aggressor row activations compared to {existing DRAM access patterns}. 
We demonstrate on a real system that {a user-level program leveraging \proposal can induce bitflips more easily compared to RowPress}.~{We analyze \omcr{2}{\proposal's} implications on existing mitigations and propose a new mitigation \omcr{2}{called} \sol.} We hope our results inspire future work to build a more comprehensive understanding of DRAM read disturbance with respect to different access patterns, and enable more robust DRAM-based computing systems.

\setcounter{version}{3}

\section*{Acknowledgments}{
We thank the anonymous reviewers of DSN 2026, HPCA 2026 and MICRO 2025 for their feedback.
We thank the SAFARI Research Group members for their constructive feedback and for providing a stimulating intellectual \omcr{3}{and scholarly} environment.
\omcr{2}{We acknowledge the generous gift funding provided by our industrial partners (especially Google, Huawei, Intel, Microsoft), which has been instrumental in enabling the research we have been conducting on read disturbance in DRAM in particular and memory systems in general\omcr{3}{~\cite{mutlu2017rowhammer,mutlu2019processing,mutlu2019rowhammer,mutlu2022modern,mutlu2023fundamentally,mutlu2023retrospective,mutlu2014research,mutlu2023retrospective-raidr,mutlu2025memory,mutlu2013memory,liu13,mutlu2023retrospective-retention,mutlu2024memory,ghose.ibmjrd19,mutlu2015main,oliveira2022accelerating,singh2021fpga,gomez2022benchmarking,oliveira2021damov}}.
This work was in part supported by a Google Security and Privacy Research Award and the Microsoft Swiss Joint Research Center.}
}

\balance
\bibliographystyle{unsrt}
\bibliography{refs}

\clearpage
\clearpage
\appendix
\nobalance

\onecolumn

\begin{landscape}
\section{{\textbf{Appendix: Summary Tables of \proposal{} Characteristics of All Tested DRAM Modules}}}
\label{sec:appendix}

Tables~\ref{tab:detailed_dram} and \ref{tab:hbm2_summary} provide detailed results for the DDR4 modules and HBM2 chips listed in Table~\ref{tab:chips}, respectively. 
For each module, we report the median and minimum \threshold{} values across all tested rows under conventional double-sided RowPress and \proposal{}. Results are grouped by OTB value (\qty{0}{ns}, \qty{300}{ns}, \qty{1200}{ns}, and \qty{7800}{ns}) and reported for two data patterns (\textit{Rowstripe0} and \textit{Rowstripe1} described in \secref{sec:method}). DDR4 DRAM chips are evaluated at both \qty{50}{\celsius} and \qty{80}{\celsius}, and HBM2 chips are evaluated at room temperature.

\begin{table}[h]
\vspace{1em}
\centering
\caption{Summary of the tested DDR4 modules and their vulnerability under double-sided RowPress and \proposal{}. We report the median and minimum \threshold{} values across different OTB values and temperatures. \threshold{} denotes the minimum number of total aggressor row activations required to induce at least one bitflip within a victim row, while OTB denotes the additional total row open time budget allocated to the two aggressor rows in double-sided patterns.}

\definecolor{rowband}{RGB}{232,232,245}
\label{tab:detailed_dram}
\scriptsize
\setlength{\tabcolsep}{2pt}
\renewcommand{\arraystretch}{1.3}
\resizebox{\linewidth}{!}{%
\begin{threeparttable}
\begin{tabular}{c c c c c c c c c| c *{6}{c} c *{6}{c}}
\toprule
\multirow{4}{*}{\makecell{\textbf{Module}}} & \multirow{4}{*}{\makecell{\textbf{DIMM Part}}} & \multirow{4}{*}{\makecell{\textbf{DRAM Part}}} & \multirow{4}{*}{\makecell{\textbf{Density}\\\textbf{(Gb)}}} & \multirow{4}{*}{\makecell{\textbf{Die-}\\\textbf{Rev.}}} & \multirow{4}{*}{\textbf{DQ}} & \multirow{4}{*}{\textbf{Ranks}} & \multirow{4}{*}{\textbf{Chips}} & \multirow{4}{*}{\makecell{\textbf{DIMM}\\\textbf{Date}}} & \multicolumn{14}{c}{\makecell{\textbf{\threshold{} @ Representative Open Time Budget (OTB)}\\\textbf{Avg. (Min.), in K (×10³)}}} \\
 &  &  &  &  &  &  &  &  & \multicolumn{7}{c}{\textbf{\qty{50}{\celsius}}} & \multicolumn{7}{c}{\textbf{\qty{80}{\celsius}}} \\
\cmidrule(lr){10-16} \cmidrule(lr){17-23}
 &  &  &  &  &  &  &  &  & \multirow{2}{*}{OTB=0} & \multicolumn{2}{c}{300 ns} & \multicolumn{2}{c}{1200 ns} & \multicolumn{2}{c}{7800 ns} & \multirow{2}{*}{OTB=0} & \multicolumn{2}{c}{300 ns} & \multicolumn{2}{c}{1200 ns} & \multicolumn{2}{c}{7800 ns} \\
\cmidrule(lr){11-12} \cmidrule(lr){13-14} \cmidrule(lr){15-16} \cmidrule(lr){18-19} \cmidrule(lr){20-21} \cmidrule(lr){22-23}
 &  &  &  &  &  &  &  &  &  & RowPress & ScaleDisturb & RowPress & ScaleDisturb & RowPress & ScaleDisturb &  & RowPress & ScaleDisturb & RowPress & ScaleDisturb & RowPress & ScaleDisturb \\
\midrule
\rowcolor{rowband} S0 & M391A2G43BB2-CWE\citemod{m391a2g43bb2} & K4AAG085WB BCWE & 16 & B & 8 & 1 & 8 & 23-15 & 33.9 (15.4) & 21.0 (12.5) & 19.4 (11.9) & 17.0 (11.1) & 16.2 (9.8) & 13.3 (9.2) & 12.8 (8.8) & 34.6 (20.3) & 21.6 (12.5) & 20.0 (11.9) & 17.6 (9.8) & 16.4 (9.6) & 13.8 (9.6) & 13.4 (9.6) \\
 S1 & M391A2G43BB2-CWE\citemod{m391a2g43bb2} & K4AAG085WB BCWE & 16 & B & 8 & 1 & 8 & 23-15 & 29.0 (14.1) & 18.3 (10.7) & 16.9 (10.2) & 15.0 (9.8) & 14.2 (9.8) & 12.8 (9.0) & 12.4 (8.6) & 30.0 (18.4) & 19.1 (12.1) & 17.7 (11.9) & 15.8 (10.0) & 15.0 (9.8) & 13.4 (9.0) & 12.9 (9.0) \\
\rowcolor{rowband} S2 & M378A2G43AB3-CWE\citemod{M378A2G43AB3-CWE} & K4AAG085WA BCWE & 16 & A & 8 & 1 & 8 & 23-02 & 37.6 (19.5) & 23.5 (14.6) & 21.8 (12.7) & 19.3 (12.5) & 18.3 (11.9) & 15.9 (10.2) & 15.4 (9.8) & 40.0 (22.3) & 24.9 (15.6) & 23.1 (14.1) & 20.4 (12.1) & 19.4 (11.3) & 16.1 (8.8) & 13.6 (5.1) \\
 S3 & M378A2G43AB3-CWE\citemod{M378A2G43AB3-CWE} & K4AAG085WA BCWE & 16 & A & 8 & 1 & 8 & N/A & 33.7 (16.4) & 21.7 (12.5) & 20.1 (12.3) & 17.8 (11.3) & 16.9 (9.8) & 14.8 (9.8) & 14.3 (9.4) & 35.3 (18.9) & 22.8 (13.7) & 21.1 (12.5) & 18.7 (12.1) & 17.8 (11.3) & 15.3 (8.8) & 13.8 (4.9) \\
\rowcolor{rowband} S4 & M378A2G43AB3-CWE\citemod{M378A2G43AB3-CWE} & K4AAG085WA BCWE & 16 & A & 8 & 1 & 8 & 23-02 & 36.6 (19.0) & 23.0 (13.7) & 21.3 (12.5) & 18.8 (10.2) & 17.8 (9.8) & 15.7 (9.4) & 15.2 (9.2) & 38.6 (20.3) & 24.3 (12.5) & 22.4 (12.5) & 20.0 (11.1) & 19.0 (10.3) & 16.2 (9.6) & 13.9 (6.6) \\
 S5 & M378A1K43DB2-CTD\citemod{M378A1K43DB2-CTD} & K4A8G085WD-BCTD & 8 & D & 8 & 1 & 8 & 21-10 & 33.1 (14.6) & 20.9 (12.1) & 19.4 (11.3) & 17.1 (11.1) & 16.2 (10.0) & 13.8 (7.8) & 12.2 (3.5) & 34.1 (18.4) & 21.7 (13.3) & 20.1 (12.5) & 17.6 (11.3) & 16.7 (11.1) & 13.6 (4.9) & 10.9 (2.5) \\
\rowcolor{rowband} S6 & M378A1K43DB2-CTD\citemod{M378A1K43DB2-CTD} & K4A8G085WD-BCTD & 8 & D & 8 & 1 & 8 & 21-10 & 31.8 (16.9) & 19.9 (11.5) & 18.3 (10.5) & 16.1 (9.8) & 15.3 (9.1) & 12.8 (5.5) & 10.5 (2.9) & 32.4 (15.2) & 20.4 (12.9) & 18.8 (11.2) & 16.6 (10.5) & 15.7 (9.4) & 12.1 (4.1) & 9.7 (2.0) \\
 S7 & F4-2400C17S-8GNT\citemod{F4-2400C17S-8GNT} & K4A4G085WF-BCTD & 4 & F & 8 & 1 & 8 & N/A & 56.4 (28.1) & 33.8 (18.8) & 31.0 (15.6) & 27.2 (14.1) & 25.8 (13.3) & 21.1 (11.9) & 17.3 (6.2) & 58.3 (26.6) & 35.0 (18.8) & 32.2 (17.2) & 27.7 (15.6) & 25.4 (13.3) & 14.8 (4.3) & 12.6 (2.3) \\
\rowcolor{rowband} H0 & HMAA4GU6AJR8N-VK\citemod{HMAA4GU6AJR8N-VK} & H5ANAG8NAJR-VKC & 16 & A & 8 & 2 & 16 & 20-03 & 71.1 (33.8) & 43.5 (21.8) & 39.7 (18.8) & 35.4 (18.8) & 33.1 (18.5) & 26.5 (12.5) & 22.3 (6.6) & 70.3 (31.9) & 42.5 (20.9) & 38.7 (19.2) & 31.8 (17.4) & 27.0 (9.6) & 16.3 (2.5) & 14.3 (1.6) \\
 H1 & HMA81GU7DJR8N-WM\citemod{HMA81GU7DJR8N-WM} & H5AN8G8NDJR-WMC & 8 & D & 8 & 1 & 8 & 19-38 & 51.8 (23.0) & 32.0 (14.1) & 29.5 (12.7) & 25.9 (12.5) & 24.3 (12.3) & 17.1 (7.8) & 13.9 (4.3) & 50.8 (23.0) & 30.7 (15.0) & 28.1 (13.3) & 21.1 (11.9) & 17.3 (7.0) & 10.4 (2.1) & 9.4 (1.2) \\
\rowcolor{rowband} H2 & HMA81GU6CJR8N-VK\citemod{HMA81GU6CJR8N-VK} & H5AN8G8NCJR-VKC & 8 & C & 8 & 1 & 8 & 21-20 & 65.3 (34.4) & 39.8 (17.4) & 36.1 (16.4) & 32.0 (15.0) & 30.1 (12.9) & 21.5 (12.7) & 20.6 (12.5) & 64.9 (29.9) & 38.7 (18.9) & 34.9 (15.6) & 27.2 (15.0) & 22.3 (10.0) & 21.8 (13.3) & 20.9 (12.5) \\
 H3 & HMA82GU7CJR8N-VK\citemod{HMA82GU7CJR8N-VK} & H5AN8G8NCJRVKC & 8 & C & 8 & 2 & 16 & N/A & 60.7 (27.3) & 37.0 (18.9) & 33.5 (18.2) & 29.7 (15.6) & 27.8 (14.8) & 22.8 (14.1) & 20.0 (9.4) & 60.3 (29.0) & 36.0 (18.9) & 32.6 (16.7) & 26.5 (15.0) & 22.4 (9.6) & 13.4 (2.5) & 11.9 (1.6) \\
\rowcolor{rowband} H4 & CMV4GX4M1A2133C15\citemod{CMV4GX4M1A2133C15} & CJR & 4 & C & 8 & 1 & 8 & N/A & 59.7 (28.1) & 36.4 (18.4) & 33.0 (15.6) & 29.4 (14.1) & 27.6 (13.3) & 22.4 (11.3) & 19.3 (6.4) & 59.4 (25.0) & 36.2 (17.6) & 32.9 (16.0) & 27.2 (12.5) & 23.3 (9.6) & 13.6 (2.5) & 11.8 (1.6) \\
 M0 & MTA18ASF4G72HZ-3G2F1Z1\citemod{MTA18ASF4G72HZ-3G2F1Z1} & MT40A2G8SA-062E:F & 16 & F & 8 & 2 & 16 & 22-37 & 19.2 (12.4) & 11.7 (6.6) & 10.6 (6.3) & 9.6 (6.2) & 8.9 (6.0) & 7.5 (4.1) & 6.2 (2.3) & 19.2 (12.3) & 11.7 (7.1) & 10.6 (6.6) & 9.5 (6.2) & 8.9 (5.7) & 5.2 (2.5) & 3.5 (1.4) \\
\rowcolor{rowband} M1 & KSM32ES8/16MF\citemod{KSM32ES8/16MF} & MT40A2G8SA-062E:F & 16 & F & 8 & 1 & 8 & 24-12 & 42.0 (11.1) & 33.1 (7.7) & 30.5 (7.0) & 28.9 (6.2) & 24.6 (5.6) & 10.9 (4.7) & 6.4 (2.5) & 39.0 (10.4) & 31.1 (6.4) & 28.6 (6.2) & 24.3 (6.2) & 16.3 (5.9) & 6.0 (2.5) & 3.4 (1.6) \\
 M2 & MTA4ATF1G64HZ-3G2E1\citemod{MTA4ATF1G64HZ-3G2E1} & MT40A1G16KD-062E:E & 16 & E & 16 & 1 & 4 & 20-46 & 20.5 (12.5) & 12.3 (7.0) & 11.3 (6.6) & 10.0 (6.1) & 9.3 (5.4) & 8.3 (4.9) & 7.8 (3.5) & 20.4 (12.5) & 12.2 (6.8) & 11.2 (6.2) & 10.0 (5.9) & 9.3 (5.7) & 8.0 (3.3) & 6.5 (1.8) \\
\rowcolor{rowband} M3 & MTA4ATF1G64HZ-3G2E1\citemod{MTA4ATF1G64HZ-3G2E1} & MT40A1G16KD-062E:E & 16 & E & 16 & 1 & 4 & 20-46 & 20.6 (12.5) & 12.2 (6.4) & 11.2 (6.2) & 9.9 (6.2) & 9.3 (5.9) & 8.2 (5.7) & 7.7 (4.9) & 20.4 (11.1) & 12.1 (7.8) & 11.2 (7.0) & 9.9 (6.5) & 9.3 (6.2) & 7.7 (4.9) & 6.0 (2.9) \\
 M4 & MTA4ATF1G64HZ-3G2B2\citemod{MTA4ATF1G64HZ-3G2B2} & MT40A1G16RC-062E:B & 16 & B & 16 & 1 & 4 & 21-26 & 53.9 (25.4) & 31.0 (15.7) & 27.8 (14.5) & 24.6 (12.5) & 23.1 (11.1) & 20.2 (9.6) & 18.5 (9.2) & 51.3 (26.6) & 29.6 (14.7) & 26.7 (12.5) & 23.7 (11.0) & 22.2 (9.6) & 18.1 (9.6) & 14.0 (7.0) \\
\rowcolor{rowband} M5 & CT16G4DFD824A.M16FE\citemod{CT16G4DFD824A.M16FE} & MT40A1G8SA-075:E & 8 & E & 8 & 2 & 16 & 23-42 & 88.1 (42.4) & 49.6 (22.7) & 44.9 (21.9) & 38.2 (20.3) & 35.8 (18.9) & 25.7 (12.2) & 21.8 (7.5) & 87.1 (37.5) & 48.8 (26.5) & 44.4 (22.3) & 36.4 (20.3) & 33.7 (19.2) & 19.5 (5.6) & 16.9 (3.1) \\
\bottomrule
\end{tabular}%
\begin{tablenotes}
\item[a] We report the date code of a DRAM module in the WW-YY format (i.e., 23-15 means the module is manufactured in the $15^{\text{th}}$ week of year 2023) as marked on the label of the module. We report ``N/A'' if no date is marked on the label of a module.
\end{tablenotes}
\end{threeparttable}
}

\end{table}

\begin{table}[h]
\vspace{2em}
\centering
\definecolor{rowband}{RGB}{232,232,245}
\caption{Summary of the tested HBM2 chips and their vulnerability under double-sided RowPress and \proposal{}.}
\label{tab:hbm2_summary}
\scriptsize
\setlength{\tabcolsep}{2pt}
\renewcommand{\arraystretch}{1.3}
\begin{threeparttable}
\begin{tabular}{c c c c c c c c c| c *{6}{c}}
\toprule
\multirow{3}{*}{\makecell{\textbf{Module}}} & \multirow{3}{*}{\makecell{\textbf{DIMM Part}}} & \multirow{3}{*}{\makecell{\textbf{DRAM Part}}} & \multirow{3}{*}{\makecell{\textbf{Density}\\\textbf{(Gb)}}} & \multirow{3}{*}{\makecell{\textbf{Die-}\\\textbf{Rev.}}} & \multirow{3}{*}{\textbf{DQ}} & \multirow{3}{*}{\textbf{Ranks}} & \multirow{3}{*}{\textbf{Chips}} & \multirow{3}{*}{\makecell{\textbf{DIMM}\\\textbf{Date}}} & \multicolumn{7}{c}{\makecell{\textbf{\threshold{} @ Representative Open Time Budget (OTB)}\\\textbf{Avg. (Min.), in K (×10³)}}} \\
 &  &  &  &  &  &  &  &  & \multirow{2}{*}{OTB=0} & \multicolumn{2}{c}{300 ns} & \multicolumn{2}{c}{1200 ns} & \multicolumn{2}{c}{7800 ns} \\
\cmidrule(lr){11-12} \cmidrule(lr){13-14} \cmidrule(lr){15-16} 
 &  &  &  &  &  &  &  &  &  & RowPress & ScaleDisturb & RowPress & ScaleDisturb & RowPress & ScaleDisturb \\
\midrule
\rowcolor{rowband} Chip0 & Unknown & Unknown & 8 & N/A & x2048 & N/A & 1 & N/A & 193.8 (74.0) & 114.2 (41.0) & 100.4 (32.0) & 74.5 (17.6) & 60.5 (9.6) & 38.0 (3.1) & 34.0 (1.6) \\
 Chip1 & Unknown & Unknown & 8 & N/A & x2048 & N/A & 1 & N/A & 174.5 (44.1) & 101.4 (36.0) & 87.3 (32.0) & 61.0 (15.1) & 49.2 (8.9) & 32.2 (2.2) & 29.2 (1.4) \\
\rowcolor{rowband} Chip2 & Unknown & Unknown & 8 & N/A & x2048 & N/A & 1 & N/A & 188.5 (66.0) & 114.0 (28.1) & 101.1 (24.2) & 78.6 (21.1) & 65.3 (13.6) & 39.1 (4.2) & 34.3 (2.1) \\
\bottomrule
\end{tabular}%
\begin{tablenotes}
\item[a] Unknown indicates that the module or chip identifier is not discernible by visual inspection of the HBM2 chip.
\item[b] A die revision of ``N/A'' means the die revision could not be identified.
\end{tablenotes}
\end{threeparttable}
\end{table}
\end{landscape}

\twocolumn

\makeatletter
\c@enumiv=0\relax
\makeatother

\bibliographystylemod{unsrt}
\bibliographymod{module_refs}

\end{document}